\documentclass[final,3p,times,authoryear]{elsarticle}

\usepackage{xcolor}
\usepackage{amsmath}
\usepackage{natbib}
\biboptions{authoryear,round}
\usepackage{mathtools}
\usepackage{physics}
\usepackage{stmaryrd} 
\usepackage{epstopdf}
\usepackage{bm}
\usepackage{graphicx,wrapfig}
\graphicspath{{../Figures/}}
\usepackage[T1]{fontenc}
\usepackage[normalem]{ulem}
\usepackage{algorithm}
\usepackage{algpseudocode}

\newcommand{\stkout}[1]{\ifmmode\text{\sout{\ensuremath{#1}}}\else\sout{#1}\fi}
\usepackage{mathrsfs}
\usepackage{amssymb}
\usepackage{nomencl}
\usepackage{etoolbox}
\usepackage{multicol}
\usepackage{caption,subcaption}
\usepackage{tabularx}
\usepackage[title]{appendix}
\newtheorem{remark}{Remark}%
\makenomenclature

\begin{document}

\begin{frontmatter}




  \title{A unified variational framework for phase-field fracture and third-medium contact in finite deformation hyperelasticity}

  \author[1]{Jaemin Kim\corref{cor1}}
  \cortext[cor1]{Corresponding Author}
  \ead{jaeminkim@changwon.ac.kr}

  \author[1]{Gukheon Kim}

  \author[1]{Sungmin Yoon}

  \author[2]{Dong-Hwa Lee\corref{cor2}}
  \cortext[cor2]{Corresponding Author}
  \ead{donghwalee@kaeri.re.kr}

  \address[1]{School of Mechanical Engineering, Changwon National University, Changwon 51140, Republic of Korea}
  \address[2]{LWR Fuel Technology Research Division, Korea Atomic Energy Research Institute, Daejeon 34057, Republic of Korea}

  \begin{abstract}
    This paper presents a unified variational framework that integrates phase-field fracture (PFF) and third-medium contact (TMC) within finite deformation hyperelasticity. The key idea is that both crack and contact are treated through regularization: the sharp crack topology is regularized into a diffuse damage field, while the discrete contact interface is regularized by a compliant fictitious medium with auxiliary fields. This strategy eliminates the need for explicit contact detection or crack tracking algorithms. The framework is validated through two-dimensional three-point bending and three-dimensional Brazilian disk test simulations, demonstrating the interplay between contact-induced stress concentration and crack nucleation/propagation. In particular, the Brazilian disk simulation naturally reproduces secondary crushing-type fracture zones near the contact regions---a phenomenon consistently observed in experiments yet inaccessible to simplified loading models. These results pave the way for predictive simulation of coupled contact--fracture phenomena without recourse to explicit interface tracking.
  \end{abstract}

\end{frontmatter}

\section*{Nomenclature}
\begin{center}
  \begin{tabularx}{\textwidth}{l@{\hspace{4mm}}X l@{\hspace{4mm}}X}
    \hline
    Symbol & Definition & Symbol & Definition \\
    \hline
    $\mathbf{X},\mathbf{x}$ & Reference/Current position vectors & $\mathbf{u}$ & Displacement field \\
    $\boldsymbol{\varphi}$ & Deformation mapping & $\mathbf{F}$ & Deformation gradient \\
    $J,\widetilde{J}$ & Jacobian determinant/Regularized $J$ & $\mathbf{C}$ & Right Cauchy--Green tensor \\
    $\mathbf{I}$ & Identity tensor & $I_1,I_3$ & Principal invariants of $\mathbf{C}$ \\
    $\boldsymbol{\varepsilon}$ & Infinitesimal strain tensor & $\varepsilon_i,\mathbf{n}_i$ & Principal strains and eigenvectors \\
    $\mathbf{P},\boldsymbol{\sigma}$ & First Piola--Kirchhoff/Cauchy stress & $\mathbf{N}$ & Unit outward normal \\
    $\mathbb{A}$ & Fourth-order elasticity tensor & $\mathbf{B},\mathbf{T}$ & Body force/Surface traction \\
    $\lambda,\mu,\kappa$ & Lam\'{e} parameter, shear and bulk moduli & $\Omega_0,\Omega_1,\Omega_2,\Omega_3$ & Reference configuration and subdomains \\
    $\Gamma$ & Crack surface & $S_u,S_t$ & Dirichlet/Neumann boundaries \\
    $\Pi$ & Total potential energy & $\Psi$ & Strain energy density \\
    $d,\ell$ & Phase-field damage and length scale & $g(d),w(d)$ & Degradation/Crack geometric functions \\
    $G_c$ & Fracture toughness & $c_w$ & Crack density normalization \\
    $k_{\ell}$ & Residual stiffness & $\Psi_1^{\pm}$ & Tensile/Compressive strain energies \\
    $\varepsilon$ & Jacobian regularization parameter & $\gamma$ & Third-medium scaling parameter \\
    $\omega,\boldsymbol{\xi}$ & Scalar/Vector microstresses & $\mathbf{W}$ & Skew-symmetric part of $\mathbf{F}$ \\
    $p,q$ & Auxiliary regularization fields & $\alpha_r$ & Gradient regularization \\
    $\beta_1,\beta_2$ & Penalty parameters & & \\
    \hline
  \end{tabularx}
\end{center}


\section{Introduction}\label{sec:intro}

Contact and fracture phenomena are widely encountered in engineering applications across multiple length scales, from nanoindentation testing for thin-film characterization \citep{saha2002effects,hay2011measuring,xiang2007measuring,kim2025film} to impact-induced delamination in laminated composites \citep{peng2019investigation,boo2026isogeometric}, interfacial failure in thermal barrier coatings \citep{di2014new,jiang2018numerical}, crack formation in battery electrodes \citep{li2011crack,chen2019approaching,boo2025multiphysics}, and rupture of biological tissues under physiological loading \citep{gultekin2016phase,fereidoonnezhad2017mechanobiological,dalbosco2024multiscale,kim2025vivo}. The simultaneous occurrence of contact and fracture is particularly prevalent in indentation-induced failure scenarios, where the concentrated stresses beneath a rigid indenter can nucleate cracks that subsequently propagate through the substrate \citep{harandi2023numerical}. Despite significant advances in computational mechanics, the unified treatment of these phenomena within a computational framework remains an open challenge because of the inherent complexity of contact algorithms themselves. Classical contact formulations require a sequence of non-trivial algorithmic steps: (i) contact detection through closest-point projection to identify active contact pairs on evolving surfaces, (ii) master--slave surface designation, which introduces asymmetry and can bias the solution, and (iii) enforcement of the non-penetration constraint. The penalty method, while straightforward to implement, introduces a user-defined stiffness parameter whose choice governs a trade-off between residual penetration and ill-conditioning; the Lagrange multiplier method enforces the constraint exactly but enlarges the algebraic system with a saddle-point structure and requires satisfaction of the inf-sup (Ladyzhenskaya--Babu\v{s}ka--Brezzi) condition for stable discretization; the augmented Lagrangian approach combines elements of both but at the cost of an iterative update loop \citep{wriggers2006computational}. This challenge is even greater when finite deformation is taken into account, as large sliding and evolving surface topology demand frequent updates of the contact search and projection operators. Coupling such contact formulations with fracture introduces a further layer of complexity. Crack propagation continuously creates new internal surfaces that may themselves come into contact---for instance, crack-face closure under compressive loading or debris interaction during fragmentation. This means the contact detection and constraint enforcement must operate on a dynamically evolving and a priori unknown topology, a requirement that is difficult to satisfy within conventional node-to-segment or segment-to-segment frameworks without substantial algorithmic augmentation.

For these reasons, numerous computational approaches have been developed to study fracture without explicit surface tracking. Early approaches based on remeshing tracked crack surfaces explicitly and required costly mesh updates at each propagation step \citep{carol1997normal}. The extended/generalized finite element method (XFEM/GFEM) alleviated this by enriching the approximation space with discontinuous and near-tip functions, enabling cracks to propagate independently of the mesh \citep{moes1999finite,belytschko2009review}; however, robust implementation in three dimensions and for multiple interacting cracks remains challenging. Cohesive zone models insert traction--separation relations along predefined element interfaces \citep{xu1994numerical,park2011cohesive}, but their accuracy depends on element alignment with the (unknown) crack path. Phase-field fracture (PFF) methods have emerged as an alternative paradigm that circumvents these difficulties, enabling the simulation of crack initiation, propagation, branching, and coalescence without explicit crack tracking \citep{francfort1998revisiting,bourdin2000numerical,miehe2010thermodynamically}. The approach regularizes the sharp crack topology through a continuous damage field $d \in [0,1]$ that interpolates between intact ($d=0$) and fully broken ($d=1$) material states. This regularization transforms the free-discontinuity problem of Griffith fracture into a coupled boundary value problem amenable to standard finite element discretization. A key advantage is that crack path and topology emerge naturally from energy minimization, eliminating the need for crack tracking algorithms \citep{borden2012phase,kristensen2021assessment}. The variational structure also provides a thermodynamically consistent framework for incorporating history-dependent irreversibility and various energy decompositions to prevent crack healing under compression \citep{amor2009regularized}. The model has been extended to the case of finite deformations in soft materials \citep{tang2019phase,ye2020large,pranavi2024unifying,kim2024finite}, anisotropic fracture \citep{wu2020anisotropic,pranavi2024anisotropic,xue2024egd}, and length-scale insensitive formulations \citep{pillai2024length}. The approach has also been applied to thin-film and film-substrate systems \citep{baldelli2014variational,guillen2019fracture,harandi2023numerical,li2024phase,kim2025film}.

On the contact mechanics, the third-medium contact (TMC) method has been developed as an elegant alternative to classical contact formulations based on penalty methods, Lagrange multipliers, or mortar approaches \citep{wriggers2025cma,wriggers2025first}. The central idea is to introduce a compliant fictitious medium---the ``third medium''---that fills the gap between potentially contacting bodies. As the bodies approach each other, the third medium compresses and naturally transmits contact forces, eliminating the need for contact detection algorithms and the associated algorithmic complexity. Notably, the resulting contact tractions have been shown to closely reproduce classical Hertzian contact solutions \citep{johnson1982one,wriggers2026third}, confirming the physical fidelity of the approach. The method is particularly attractive for large-deformation problems where contact surfaces undergo significant changes in geometry. However, the severe compression of the third medium can lead to mesh distortion and element inversion. To address this issue, \citet{dahlberg2025rotation} proposed an auxiliary-field regularization, in which additional fields track local rotation and volumetric deformation to maintain mesh quality. Extensions to thermo-mechanical coupling \citep{wriggers2026third} and three-dimensional hyperelastic contact \citep{xu2025three} further demonstrate the versatility of the TMC framework.

Despite these parallel advances in PFF and TMC, a unified formulation that couples both methods within a single variational framework has not yet been established; as a result, a deep understanding of the interplay between contact and fracture through computational simulation is still lacking. While these two methods have been developed independently, it turns out that a natural synergy exists between them: PFF regularizes the sharp crack topology into a diffuse damage field governed by a length-scale parameter $\ell$, while TMC regularizes the discrete contact interface through a compliant fictitious medium controlled by a scaling parameter $\gamma$. This should not come as a surprise, since both methods fundamentally rely on the same regularization strategy---replacing discontinuous surfaces with continuous field descriptions. The above arguments lead to the conclusion that both phenomena can be naturally unified within a single energy functional. Combining them is further attractive because the third medium naturally handles the evolving contact topology created by crack propagation---new crack faces that come into contact are automatically resolved through the fictitious medium without additional algorithmic treatment. The principal challenge is the construction of a single energy functional that consistently combines: (i) a phase-field--degraded substrate with a tensile-compressive energy split, (ii) stiff contacting bodies, (iii) a highly compliant third medium undergoing extreme compression ($J\to 0$), and (iv) auxiliary regularization fields that maintain mesh quality under severe deformation.

The objective of this work is to develop such a unified variational framework coupling PFF and TMC within a thermodynamically consistent finite deformation hyperelastic setting. This work does not introduce new constitutive models for fracture or contact; instead, it explores the unification of existing PFF and TMC methods into a single potential energy functional, enabling the simulation of contact-induced fracture without explicit contact detection or algorithmic modifications to either constituent method. The present work limits itself to employing the AT2 crack density functional and frictionless Neo-Hookean hyperelasticity, constructing a unified energy functional that encompasses phase-field fracture in the substrate, Neo-Hookean hyperelasticity in the contacting bodies, and third-medium contact with auxiliary-field regularization. A spectral strain energy decomposition \citep{miehe2010thermodynamically} separates principal tensile and compressive contributions, preventing crack healing under compression. A regularized Jacobian treatment ensures robust constitutive evaluation in regions of extreme deformation where $J \to 0$. The framework is validated through two-dimensional three-point bending and three-dimensional Brazilian disk test simulations that demonstrate the interplay between contact stress concentration and damage evolution. Notably, the Brazilian disk simulation captures secondary crushing-type fracture near the evolving contact regions, a feature that is consistent with experimental observations but cannot be reproduced by models employing prescribed contact geometries.

The remainder of this paper is organized as follows. Section~\ref{sec:unified} establishes the kinematic framework for finite deformations, the domain decomposition into substrate, indenter, and third-medium subdomains, and the total potential energy functional. Section~\ref{sec:general-theory} develops the phase-field fracture theory, including the regularization of the crack topology, balance principles, thermodynamic constitutive relations, and the strain energy decomposition. Section~\ref{sec:tmc} specifies the third-medium contact theory, comprising the indenter model and the auxiliary-field regularization for mesh quality control. Section~\ref{sec:weak} derives the weak form of the governing equations, presents the finite element discretization, and describes the staggered solution algorithm. Section~\ref{sec:examples} presents numerical examples, beginning with independent validation of the TMC and PFF components, followed by their combined application to a three-point bending test and a three-dimensional Brazilian disk test. Section~\ref{sec:conclusions} concludes the paper with remarks on future extensions. Appendix~\ref{sec:appendix_thermo} provides the thermodynamic derivation of the constitutive relations for phase-field damage, Appendix~\ref{sec:appendix_crack} derives the crack evolution equation, and Appendix~\ref{sec:appendix_stress} presents the stress tensor expressions for each subdomain.

\begin{remark}[Notation]\label{rmk:notation}
  Throughout this work, the following conventions are adopted:
  \begin{enumerate}
    \item Vectors and second-order tensors are denoted by bold symbols (e.g., $\mathbf{u}$, $\mathbf{F}$, $\mathbf{P}$), while scalars are set in italic (e.g., $d$, $J$, $\ell$).
    \item The gradient operator $\nabla$ denotes differentiation with respect to material coordinates $\mathbf{X}$ unless otherwise specified. For clarity, $\nabla_{\mathbf{X}}(\cdot)$ is occasionally used.
    \item The dyadic product is denoted by $\otimes$, the dot product by $\cdot$, and the double contraction by $:$, e.g., $\mathbf{A}:\mathbf{B} = A_{ij}B_{ij}$.
    \item For a second-order tensor $\mathbf{A}$, $\mathrm{tr}(\mathbf{A})$ and $\det(\mathbf{A})$ denote its trace and determinant, and $\mathbf{A}^{-\mathrm{T}}$ denotes the inverse transpose.
    \item The Einstein summation convention is adopted: repeated indices imply summation. Unless noted otherwise, Latin indices $i,j,k,\ldots$ take values in $\{1,2,3\}$ (with plane strain enforced by setting $F_{33}=1$).
    \item The identity tensor is denoted by $\mathbf{I}$. The deformation gradient is $\mathbf{F}=\mathbf{I}+\nabla\mathbf{u}$ and the Jacobian is $J=\det\mathbf{F}$.
    \item Boundary sets are denoted by $S_u$ (Dirichlet) and $S_t$ (Neumann). Prescribed values are indicated by a check accent, e.g., $\check{\mathbf{u}}$ and $\check{d}$.
    \item When the domain is decomposed into subdomains, integrals are always taken over the appropriate region: fracture terms are restricted to $\Omega_2$, while auxiliary-field regularization terms are restricted to $\Omega_3$.
  \end{enumerate}
\end{remark}

\section{A unified framework}\label{sec:unified}

\subsection{Kinematics}

Consider a body occupying the reference configuration $\Omega_0 \subset \mathbb{R}^{n_{\mathrm{dim}}}$ with $n_{\mathrm{dim}} \in \{2,3\}$. The deformation is described by the mapping $\boldsymbol{\varphi}: \Omega_0 \times [0,T] \to \mathbb{R}^{n_{\mathrm{dim}}}$ such that
\begin{align}
  \mathbf{x} = \boldsymbol{\varphi}(\mathbf{X},t),
\end{align}
where $\mathbf{X}$ and $\mathbf{x}$ denote material and spatial positions, respectively. The displacement field is $\mathbf{u} = \mathbf{x} - \mathbf{X}$, and the deformation gradient tensor is
\begin{align}
  \mathbf{F} = \nabla_{\mathbf{X}}\boldsymbol{\varphi} = \mathbf{I} + \nabla_{\mathbf{X}}\mathbf{u}, \label{eqn:F}
\end{align}
with Jacobian determinant $J = \det\mathbf{F} > 0$ for admissible (orientation-preserving) deformations. The right Cauchy--Green deformation tensor and its principal invariants are
\begin{align}
  \mathbf{C} = \mathbf{F}^{\mathrm{T}}\mathbf{F}, \qquad I_1 = \mathrm{tr}(\mathbf{C}), \qquad I_3 = \det(\mathbf{C}) = J^2.
\end{align}
For isotropic hyperelastic materials, the strain energy density depends on $\mathbf{F}$ only through the invariants $I_1$, $I_2 = \frac{1}{2}[({\rm tr}\,\mathbf{C})^2 - {\rm tr}(\mathbf{C}^2)]$, and $I_3$ (or equivalently $J$).

\subsection{Domain decomposition}

The computational domain is decomposed into three non-overlapping subdomains:
\begin{align}
  \Omega_0 = \Omega_1 \cup \Omega_2 \cup \Omega_3, \qquad \Omega_1 \cap \Omega_2 = \Omega_1 \cap \Omega_3 = \Omega_2 \cap \Omega_3 = \emptyset, \label{eqn:domain}
\end{align}
where $\Omega_1$ is the substrate domain where fracture may evolve, typically a deformable solid modeled with phase-field damage; $\Omega_2$ is the indenter or contact body domain, typically modeled as a stiff elastic solid without damage evolution; and $\Omega_3$ is the third-medium domain occupying the gap between $\Omega_1$ and $\Omega_2$, which is a compliant fictitious material that facilitates contact enforcement.

\begin{figure}[t]
  \centering
  \includegraphics[width=0.9\textwidth]{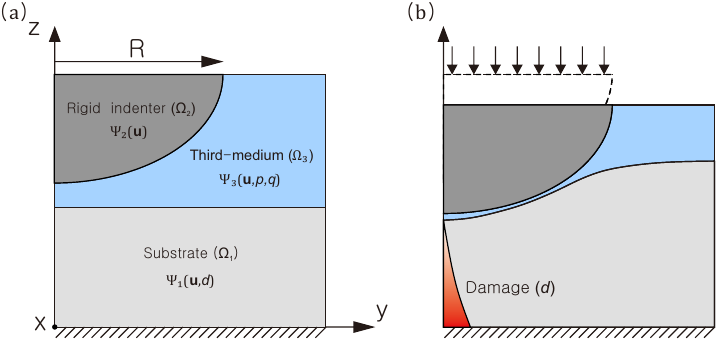}
  \caption{Schematic drawing for the unified PFF-TMC framework. (a) Reference configuration showing the substrate $\Omega_1$, indenter $\Omega_2$, and third-medium domain $\Omega_3$ with their respective field variables. (b) Deformed configuration after the mapping $\boldsymbol{\varphi}$, illustrating the compressed third medium ($J \to 0$), contact pressure transmission, stress concentration beneath the indenter, and phase-field damage evolution ($d \to 1$) in the substrate.}
  \label{fig:kinematics}
\end{figure}

The displacement field $\mathbf{u}$ is defined globally on $\Omega_0$ with required regularity across subdomain interfaces. The phase-field variable $d \in [0,1]$ is defined on $\Omega_1$ only, while the auxiliary regularization fields $p$ and $q$ are defined on $\Omega_3$ only. The reference and deformed configurations, along with the domain decomposition, are illustrated in Fig.~\ref{fig:kinematics}.

\subsection{Total potential energy}

The coupled contact-fracture problem is formulated through the minimization of a total potential energy functional. For quasi-static problems, this functional takes the general form
\begin{align}
  \Pi(\mathbf{u},d,p,q) = \Pi_{\mathrm{int}}(\mathbf{u},d,p,q) + \Pi_{\mathrm{frac}}(d) - \Pi_{\mathrm{ext}}(\mathbf{u}), \label{eqn:Pi_general}
\end{align}
where $\Pi_{\mathrm{int}}$ is the internal elastic energy, $\Pi_{\mathrm{frac}}$ is the fracture dissipation energy, and $\Pi_{\mathrm{ext}}$ is the external work.
The external work due to body forces $\mathbf{B}$ and surface tractions $\mathbf{T}$ on the Neumann boundary $S_t$ is
\begin{align}
  \Pi_{\mathrm{ext}}(\mathbf{u}) = \int_{\Omega_0} \mathbf{B}\cdot\mathbf{u}\,\mathrm{d}V + \int_{S_t} \mathbf{T}\cdot\mathbf{u}\,\mathrm{d}A.
\end{align}
The fracture dissipation energy accounts for the creation of new crack surfaces (Griffith form) \citep{francfort1998revisiting,miehe2010thermodynamically,borden2012phase,wu2017unified,vajari2023investigation,kim2025film}:
\begin{align}
  \Pi_{\mathrm{frac}} = \int_{\Gamma} G_c\,\mathrm{d}S, \label{eqn:Pi_frac}
\end{align}
where $\Gamma$ denotes the (sharp) crack surface. In the phase-field formulation, this surface term is approximated by the volumetric crack density functional in Eq.~\eqref{eqn:Gamma_approx}.
The internal elastic energy is the sum of contributions from each subdomain:
\begin{align}
  \Pi_{\mathrm{int}}(\mathbf{u},d,p,q) = \int_{\Omega_1} \Psi_1(\mathbf{F},d)\,\mathrm{d}V + \int_{\Omega_2} \Psi_2(\mathbf{F})\,\mathrm{d}V + \int_{\Omega_3} \Psi_3(\mathbf{F},p,q)\,\mathrm{d}V, \label{eqn:Pi_int}
\end{align}
where $\Psi_1$, $\Psi_2$, and $\Psi_3$ are the strain energy densities for the substrate, indenter, and third medium, respectively.

\section{A phase-field fracture theory}\label{sec:general-theory}

\subsection{Phase-field regularization of fracture}

\begin{figure}[t]
  \centering
  \includegraphics[width=0.9\textwidth]{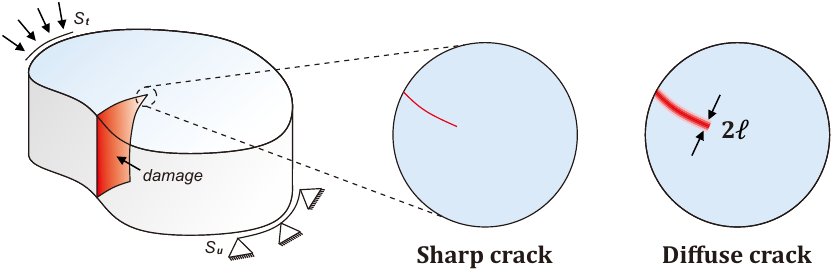}
  \caption{Phase-field regularization of fracture. A body subjected to boundary conditions ($S_u$: Dirichlet, $S_t$: Neumann) containing a sharp crack (center) is approximated by a diffuse crack band (right) with characteristic width $2\ell$, where the damage field $d$ transitions continuously from $d=0$ (intact) to $d=1$ (fully broken).}
  \label{fig:pff_concept}
\end{figure}

Following the variational approach to brittle fracture \citep{francfort1998revisiting,miehe2010thermodynamically,borden2012phase,wu2017unified,vajari2023investigation,kim2025film,bourdin2000numerical}, the sharp crack is regularized by a continuous phase-field variable $d \in [0,1]$, where $d=0$ represents intact material and $d=1$ represents fully broken material. The crack surface energy is approximated by the crack density functional
\begin{align}
  \Pi_{\mathrm{frac}} = \int_{\Gamma} G_c \,\mathrm{d}S \approx \int_{\Omega_1} \frac{G_c}{c_w}\left(\frac{w(d)}{\ell} + \ell|\nabla d|^2\right)\mathrm{d}V, \label{eqn:Gamma_approx}
\end{align}
where $G_c$ is the fracture toughness, $\ell$ is the regularization length scale, $w(d)$ is the crack geometric function, and $c_w$ is a normalization constant. This regularization is illustrated schematically in Fig.~\ref{fig:pff_concept}, where the sharp crack surface $\Gamma$ is replaced by a diffuse damage band of characteristic width $2\ell$.

\subsection{Balance principles}\label{sec:balance}

The governing equations consist of two balance principles: the macroscopic mechanical equilibrium and the microscopic crack evolution.
The macroscopic balance is governed by the balance of linear momentum in the reference configuration:
\begin{alignat}{2}
  \nabla\cdot\mathbf{P} + \mathbf{B} &= 0 \quad &&\text{in } \Omega_0, \label{eqn:macro_eq} \\
  \mathbf{u} &= \check{\mathbf{u}} \quad &&\text{on } S_u, \label{eqn:macro_bc_u} \\
  \mathbf{P}\cdot\mathbf{N} &= \mathbf{T} \quad &&\text{on } S_t, \label{eqn:macro_bc_t}
\end{alignat}
where $\mathbf{B}$ is the body force per unit reference volume, $\check{\mathbf{u}}$ is the prescribed displacement on the Dirichlet boundary $S_u$, and $\mathbf{T}$ is the prescribed traction on the Neumann boundary $S_t$.

The microscopic balance governing crack evolution involves the scalar microstress $\omega$ and vector microstress $\boldsymbol{\xi}$ \citep{francfort1998revisiting,miehe2010thermodynamically,borden2012phase,wu2017unified,vajari2023investigation,kim2025film,mao2018theory}:
\begin{alignat}{2}
  \nabla\cdot\boldsymbol{\xi} - \omega &= 0 \quad &&\text{in } \Omega_1, \label{eqn:micro_eq} \\
  d &= \check{d} \quad &&\text{on } S_d. \label{eqn:micro_bc}
\end{alignat}
where $\omega$ is the scalar microstress conjugate to the damage variable $d$, $\boldsymbol{\xi}$ is the vector microstress conjugate to the damage gradient $\nabla d$, $\check{d}$ is the prescribed damage on the boundary $S_d$.

\subsection{Constitutive relations}\label{sec:thermo}
The constitutive framework for phase-field fracture is derived from thermodynamic principles following the Coleman--Noll procedure \citep{coleman1963thermodynamics}. For isothermal processes, a Helmholtz free energy density is postulated as $\Psi = \Psi(\mathbf{F}, d, \nabla d)$ that decomposes into elastic and fracture contributions:
\begin{align}
  \Psi(\mathbf{F}, d, \nabla d) = \Psi_{\mathrm{el}}(\mathbf{F}, d) + \Psi_{\mathrm{frac}}(d, \nabla d), \label{eqn:Psi_decomp}
\end{align}
where
\begin{align}
  \Psi_{\mathrm{el}}(\mathbf{F}, d) &= g(d)\,\Psi_{\mathrm{und}}(\mathbf{F}), \label{eqn:Psi_el}\\
  \Psi_{\mathrm{frac}}(d, \nabla d) &= \frac{G_c}{c_w}\left(\frac{w(d)}{\ell} + \ell|\nabla d|^2\right). \label{eqn:Psi_frac_thermo}
\end{align}
Here, $g(d)$ is the degradation function, $\Psi_{\mathrm{und}}(\mathbf{F})$ is the undamaged strain energy density, and $\Psi_{\mathrm{frac}}$ is the fracture surface energy density consistent with Eq.~\eqref{eqn:Gamma_approx}.
The specific functional forms of $g(d)$ and $w(d)$ are deferred to Section~\ref{sec:specific_choice} to keep the derivation general.

The second law of thermodynamics, through the Coleman--Noll procedure (see Appendix~\ref{sec:appendix_thermo} for detailed derivation), yields the constitutive relations:
\begin{align}
  \mathbf{P} &= \frac{\partial \Psi}{\partial \mathbf{F}} = g(d)\frac{\partial \Psi_{\mathrm{und}}(\mathbf{F})}{\partial \mathbf{F}}, \label{eqn:P_constitutive} \\
  \boldsymbol{\xi} &= \frac{\partial \Psi}{\partial \nabla d} = \frac{2G_c\ell}{c_w}\nabla d, \label{eqn:xi_constitutive} \\
  \omega &= \frac{G_c}{c_w\ell}w'(d) + g'(d)\,\Psi_{\mathrm{und}}(\mathbf{F}). \label{eqn:omega_constitutive}
\end{align}

\subsection{Strain energy density for the substrate}\label{sec:substrate_energy}

For substrates composed of stiff, brittle materials where strains remain small prior to fracture, a linearized (small-strain) formulation is employed. The infinitesimal strain tensor is
\begin{align}
  \boldsymbol{\varepsilon} = \mathrm{sym}(\nabla\mathbf{u}) = \frac{1}{2}\left(\nabla\mathbf{u} + (\nabla\mathbf{u})^{\mathrm{T}}\right), \label{eqn:eps_def}
\end{align}
and the isotropic linear elastic strain energy density takes the form
\begin{align}
  \Psi_{\mathrm{und}}(\boldsymbol{\varepsilon}) = \frac{\lambda}{2}(\mathrm{tr}\,\boldsymbol{\varepsilon})^2 + \mu\,\mathrm{tr}(\boldsymbol{\varepsilon}^2), \label{eqn:Psi_und_smallstrain}
\end{align}
where $\lambda = \kappa - 2\mu/3$ is the first Lam\'{e} parameter, $\mu$ is the shear modulus, and $\kappa$ is the bulk modulus.

\begin{remark}[Mixed large/small-strain formulation]\label{rmk:mixed_strain}
  The substrate (fracturing domain $\Omega_1$) employs the small strain approximation \eqref{eqn:eps_def}--\eqref{eqn:Psi_und_smallstrain}, which is appropriate for stiff, brittle materials where strains remain small until failure. The third medium ($\Omega_3$) retains the large deformation Neo-Hookean formulation, as the third medium undergoes severe compression ($J \to 0$) during contact and requires a large deformation formulation.
\end{remark}

\subsection{Spectral strain energy decomposition}\label{sec:substrate_energy_split}

To preclude spurious damage evolution under compressive stress states, the spectral decomposition proposed by \citet{miehe2010thermodynamically} is adopted. The strain tensor is decomposed into its principal components:
\begin{align}
  \boldsymbol{\varepsilon} = \sum_{i=1}^{n_{\mathrm{dim}}} \varepsilon_i \, \mathbf{n}_i \otimes \mathbf{n}_i, \label{eqn:eps_spectral}
\end{align}
where $\varepsilon_i$ are the principal strains (eigenvalues of $\boldsymbol{\varepsilon}$) and $\mathbf{n}_i$ the corresponding eigenvectors. The positive (tensile) and negative (compressive) parts are defined using the Macaulay brackets $\langle x \rangle_{\pm}$:
\begin{align}
  \boldsymbol{\varepsilon}^{+} = \sum_{i=1}^{n_{\mathrm{dim}}} \langle\varepsilon_i\rangle_{+} \, \mathbf{n}_i \otimes \mathbf{n}_i, \qquad
  \boldsymbol{\varepsilon}^{-} = \sum_{i=1}^{n_{\mathrm{dim}}} \langle\varepsilon_i\rangle_{-} \, \mathbf{n}_i \otimes \mathbf{n}_i, \label{eqn:eps_pm}
\end{align}
where $\langle x \rangle_{+} = \max(x,0)$ and $\langle x \rangle_{-} = \min(x,0)$, so that $\boldsymbol{\varepsilon} = \boldsymbol{\varepsilon}^{+} + \boldsymbol{\varepsilon}^{-}$.

The tensile and compressive strain energies are then defined as:
\begin{align}
  \Psi_1^{+}(\boldsymbol{\varepsilon}) &= \frac{\lambda}{2}\langle\mathrm{tr}\,\boldsymbol{\varepsilon}\rangle_{+}^{2} + \mu\sum_{i=1}^{n_{\mathrm{dim}}} \langle\varepsilon_i\rangle_{+}^{2}, \label{eqn:Psi_plus} \\
  \Psi_1^{-}(\boldsymbol{\varepsilon}) &= \frac{\lambda}{2}\langle\mathrm{tr}\,\boldsymbol{\varepsilon}\rangle_{-}^{2} + \mu\sum_{i=1}^{n_{\mathrm{dim}}} \langle\varepsilon_i\rangle_{-}^{2}. \label{eqn:Psi_minus}
\end{align}
The first term in each expression governs the volumetric response, where $\langle\mathrm{tr}\,\boldsymbol{\varepsilon}\rangle_{+}$ captures dilatation and $\langle\mathrm{tr}\,\boldsymbol{\varepsilon}\rangle_{-}$ captures contraction. The second term governs the deviatoric response, where only the tensile principal strains contribute to $\Psi_1^{+}$ and only the compressive principal strains contribute to $\Psi_1^{-}$. This decomposition ensures that: (i) tensile stresses drive damage evolution through $\Psi_1^{+}$, and (ii) fully damaged material ($d=1$) retains compressive stiffness through $\Psi_1^{-}$, preventing unphysical interpenetration of crack faces.

The degraded strain energy density for the substrate incorporates the phase-field damage:
\begin{align}
  \Psi_1(\boldsymbol{\varepsilon},d) = (g(d) + k_{\ell})\Psi_1^{+}(\boldsymbol{\varepsilon}) + \Psi_1^{-}(\boldsymbol{\varepsilon}), \label{eqn:Psi_s}
\end{align}
where $k_{\ell} \ll 1$ (typically $10^{-6}$) is a small residual stiffness parameter that ensures positive definiteness of the tangent stiffness matrix in fully damaged regions ($d \to 1$). The specific choice of $g(d)$ is given in Section~\ref{sec:specific_choice}.

With this decomposition, the constitutive relations become:
\begin{align}
  \boldsymbol{\sigma} &= (g(d) + k_{\ell})\frac{\partial \Psi_1^{+}}{\partial \boldsymbol{\varepsilon}} + \frac{\partial \Psi_1^{-}}{\partial \boldsymbol{\varepsilon}}, \label{eqn:P_split} \\
  \omega &= \frac{G_c}{c_w\ell}w'(d) + g'(d)\,\Psi_1^{+}(\boldsymbol{\varepsilon}), \label{eqn:omega_split}
\end{align}
where $\boldsymbol{\sigma}$ is the Cauchy stress tensor in the substrate. The detailed derivation of the crack evolution equation~\eqref{eqn:omega_split} is provided in Appendix~\ref{sec:appendix_crack}.

\begin{remark}[Eigenvalue computation in 2D and 3D]\label{rmk:eigenvalues}
  In two dimensions (plane strain), the principal strains are obtained analytically from the quadratic formula for the $2\times 2$ strain tensor, with $\varepsilon_3 = 0$. In three dimensions, the three principal strains are computed using Cardano's formula for the characteristic cubic of the $3\times 3$ strain tensor \citep{itskov2007tensor,ang2022stabilized}. Given the strain invariants
  \begin{align}
    I_1 = \mathrm{tr}\,\boldsymbol{\varepsilon}, \qquad
    I_2 = \tfrac{1}{2}\bigl[(\mathrm{tr}\,\boldsymbol{\varepsilon})^2 - \mathrm{tr}\,\boldsymbol{\varepsilon}^2\bigr], \qquad
    I_3 = \det\boldsymbol{\varepsilon},
  \end{align}
  the depressed-cubic parameters are
  \begin{align}
    p = I_2 - \frac{I_1^2}{3}, \qquad
    q = \frac{-2I_1^3 + 9I_1 I_2 - 27I_3}{27},
  \end{align}
  and the three eigenvalues follow from the trigonometric solution
  \begin{align}
    m = \sqrt{-\frac{p}{3}}, \qquad
    \theta = \frac{1}{3}\arccos\,\left(\frac{-q}{2m^3}\right), \qquad
    \varepsilon_i = 2m\cos\,\left(\theta - \frac{2(i-1)\pi}{3}\right) + \frac{I_1}{3}.
  \end{align}
  In the numerical implementation, smooth approximations $\langle x \rangle_{+} \approx (x + \sqrt{x^2 + \delta^2})/2$ with $\delta = 10^{-5}$ replace the Macaulay brackets, and additive regularization replaces conditional floor operations in the Cardano formula, to ensure well-defined Jacobians for Newton's method.
\end{remark}

\subsection{Specific choices of degradation and crack geometric functions}\label{sec:specific_choice}
To close the formulation, the crack density and degradation functions are now specified. Two commonly used models are the AT1 and AT2 families \citep{pham2011gradient,ambrosio1990approximation,bourdin2000numerical}. The AT1 model exhibits a sharp onset of damage with an elastic threshold, whereas the AT2 model lacks an elastic threshold and yields a smooth ($C^1$-continuous) energy landscape with a gradual transition from intact to damaged states. The AT2 model is adopted in this work for the following reasons: (i) it provides better numerical stability due to its smooth energy landscape, (ii) it is well-suited for quasi-static simulations where gradual damage evolution is expected. Accordingly, the specific forms are
\begin{align}
  w(d) = d^2, \qquad c_w = 2. \label{eqn:AT2_choice}
\end{align}

A degradation function $g(d)$ monotonically diminishes the effective elastic stiffness with increasing damage. The quadratic form is adopted, and thermodynamic irreversibility is enforced through
\begin{align}
  g(d) = (1-d)^2, \qquad \dot{d} \geq 0, \label{eqn:g_irreversibility}
\end{align}
where the irreversibility constraint is enforced algorithmically by maintaining a lower bound on the damage field (see Appendix~\ref{sec:appendix_crack}).

\section{A third-medium contact theory}\label{sec:tmc}

This section presents the third-medium contact (TMC) formulation, which enables contact enforcement through a compliant fictitious medium without explicit contact detection algorithms.

\subsection{Third-medium strain energy density}

The third medium is a compliant fictitious material that fills the gap between contacting bodies, enabling contact enforcement without explicit contact detection algorithms. As the contacting bodies approach each other, the third medium compresses and naturally transmits contact forces through its constitutive response. Following \citet{wriggers2025cma,dahlberg2025rotation}, a multi-component energy density is employed with auxiliary-field regularization to maintain mesh quality under severe deformation.

The total third-medium energy density consists of elastic and regularization contributions:
\begin{align}
  \Psi_3(\mathbf{F},p,q) = \Psi_{3}^{\mu}(\mathbf{F}) + \Psi_{3}^{\gamma}(\mathbf{F}) + \Psi_{3}^{\mathrm{reg}}(\mathbf{F},p,q). \label{eqn:Psi_m}
\end{align}
The first two terms constitute the elastic response, while the third term provides auxiliary-field regularization. Each component is defined below.

The base isochoric term $\Psi_{3}^{\mu}$ governs the deviatoric (volume-preserving) response:
\begin{align}
  \Psi_{3}^{\mu}(\mathbf{F}) = \frac{\mu_3}{2}(J^{-2/3}I_1 - 3), \label{eqn:Psi_mmu}
\end{align}
and the $\gamma$-scaled term $\Psi_{3}^{\gamma}$ supplements the full Neo-Hookean response at a reduced stiffness level:
\begin{align}
  \Psi_{3}^{\gamma}(\mathbf{F}) = \gamma\left[\frac{\kappa_3}{2}(\ln J)^2 + \frac{\mu_3}{2}(J^{-2/3}I_1 - 3)\right], \qquad 0 < \gamma \ll 1. \label{eqn:Psi_mg}
\end{align}
The small scaling parameter $\gamma$ (typically $10^{-6}$ to $10^{-9}$) ensures that the third medium is orders of magnitude more compliant than the surrounding solids, while the volumetric term $(\ln J)^2$ provides the contact pressure that prevents interpenetration as the gap closes (see Remark~\ref{rmk:J_safe}). Under severe compression, the third-medium mesh can experience extreme distortion; to control mesh quality, the regularization term $\Psi_{3}^{\mathrm{reg}}$ regularizes local rotation and volumetric deformation through auxiliary fields \citep{dahlberg2025rotation,xu2025three}.

\subsection{Auxiliary-field regularization}
In three dimensions, the local rotation is characterized by the skew-symmetric part of the deformation gradient:
\begin{align}
  \mathbf{W} = \frac{1}{2}(\mathbf{F} - \mathbf{F}^{\mathrm{T}}), \label{eqn:W_skew}
\end{align}
which has three independent components $W_{12}$, $W_{13}$, and $W_{23}$. Three auxiliary fields $\mathbf{p} = (p_1, p_2, p_3)$ are introduced to approximate these rotation components. The regularization energy is:
\begin{align}
  \Psi_{3}^{\mathrm{reg}}(\mathbf{F},\mathbf{p},q) = \frac{\gamma}{2}\left[\sum_{i=1}^{3}\left(\beta_1(W_i - p_i)^2 + \alpha_r|\nabla p_i|^2\right) + \beta_2(J - q)^2 + \alpha_r|\nabla q|^2\right], \label{eqn:Psi_reg}
\end{align}
where $W_1 = F_{12} - F_{21}$, $W_2 = F_{13} - F_{31}$, $W_3 = F_{23} - F_{32}$ are the independent skew-symmetric components. Here, $\mathbf{p} = (p_1, p_2, p_3)$ are auxiliary fields that approximate the local rotation components, where the gradient penalties suppress spatial oscillations in the rotation field that would otherwise lead to element distortion and inversion. Similarly, $q$ is an auxiliary field approximating the local Jacobian $J$, and the associated gradient penalty suppresses volumetric deformation jumps. The penalty parameters $\beta_1$ and $\beta_2$ enforce the approximations; large values make the auxiliary fields closely track the kinematic quantities. The gradient regularization parameter $\alpha_r$ controls the smoothness of the auxiliary fields.

For plane strain (2D), a single rotation measure is sufficient. A commonly used choice is
\begin{align}
  \tan\varphi = \frac{F_{12}-F_{21}}{F_{11}+F_{22}}. \label{eqn:tan_phi}
\end{align}
An auxiliary field $p$ approximates $\tan\varphi$. The regularization reduces to
\begin{align}
  \Psi_{3,2D}^{\mathrm{reg}} = \frac{\gamma}{2}\left[\beta_1(\tan\varphi-p)^2 + \alpha_r|\nabla p|^2 + \beta_2(J-q)^2 + \alpha_r|\nabla q|^2\right].
\end{align}
In the plane-strain (2D) setting, the vector $\mathbf{p}$ reduces to a scalar field $p$, and accordingly the auxiliary fields $p$ and $q$ are both scalar. For three-dimensional problems, the full three-component vector $\mathbf{p} = (p_1, p_2, p_3)$ is retained, as described in Eq.~\eqref{eqn:Psi_reg}.

\section{Finite element method}\label{sec:weak}

\subsection{Total potential energy}

The equilibrium state is obtained by minimizing the total potential energy functional with respect to $(\mathbf{u},d,\mathbf{p},q)$:
\begin{align}
  \Pi(\mathbf{u},d,\mathbf{p},q) &= \underbrace{\int_{\Omega_1} \left[(g(d)+k_{\ell})\Psi_1^{+} + \Psi_1^{-}\right]\mathrm{d}V}_{\text{Substrate elastic energy}} + \underbrace{\int_{\Omega_1} \frac{G_c}{c_w}\left(\frac{w(d)}{\ell} + \ell|\nabla d|^2\right)\mathrm{d}V}_{\text{Substrate fracture energy}} \nonumber \\  &\quad + \underbrace{\int_{\Omega_2} \Psi_2\,\mathrm{d}V}_{\text{Rigid indenter energy}} + \underbrace{\int_{\Omega_3} \left[\Psi_{3}^{\mu} + \Psi_{3}^{\gamma} + \Psi_{3}^{\mathrm{reg}}\right]\mathrm{d}V}_{\text{Third-medium energy}} - \Pi_{\mathrm{ext}}. \label{eqn:Pi_total}
\end{align}

\subsection{Weak form}

The weak form of the problem is obtained by multiplying the strong-form equations with appropriate test functions $\delta\mathbf{u}$, $\delta d$, $\delta\mathbf{p}$, $\delta q$ and integrating over the respective domains. The momentum balance yields
\begin{align}
  \int_{\Omega_0} \mathbf{P} : \nabla\delta\mathbf{u}\,\mathrm{d}V - \int_{S_t} \mathbf{T}\cdot\delta\mathbf{u}\,\mathrm{d}A = 0, \label{eqn:weak_u}
\end{align}
where $\mathbf{P} = \partial\Psi/\partial\mathbf{F}$ is subdomain-dependent. The auxiliary field equations for the third medium are
\begin{align}
  \sum_{i=1}^{3}\int_{\Omega_3} \gamma\left[\beta_1(p_i - W_i)\delta p_i + \alpha_r\nabla p_i\cdot\nabla\delta p_i\right]\mathrm{d}V &= 0, \label{eqn:weak_p} \\
  \int_{\Omega_3} \gamma\left[\beta_2(q - J)\delta q + \alpha_r\nabla q\cdot\nabla\delta q\right]\mathrm{d}V &= 0, \label{eqn:weak_q}
\end{align}
where $W_i$ are the independent skew-symmetric components of $\mathbf{F}$ defined in Eq.~\eqref{eqn:Psi_reg}.
The phase-field evolution is governed by the variational inequality
\begin{align}
  \int_{\Omega_1}\left[(g'(d)\Psi_1^{+} + \frac{G_c}{c_w\ell}w'(d))\delta d + \frac{2G_c\ell}{c_w}\nabla d\cdot\nabla\delta d\right]\mathrm{d}V \geq 0, \label{eqn:weak_d}
\end{align}
subject to the irreversibility constraint $\dot{d} \geq 0$.

The statement of the weak form is to find the trial functions $(\mathbf{u}, d, \mathbf{p}, q) \in \mathcal{S}_u \times \mathcal{S}_d \times \mathcal{S}_p \times \mathcal{S}_q$ such that Eqs.~\eqref{eqn:weak_u}--\eqref{eqn:weak_d} are satisfied for all admissible test functions $(\delta\mathbf{u}, \delta d, \delta\mathbf{p}, \delta q) \in \mathcal{V}_u \times \mathcal{V}_d \times \mathcal{V}_p \times \mathcal{V}_q$. The sets of admissible trial functions are denoted by
\begin{align}
  \mathcal{S}_u &= \{\mathbf{u} \,|\, \mathbf{u} \in [H^1(\Omega_0)]^{n_{\mathrm{dim}}},\; \mathbf{u} = \check{\mathbf{u}} \text{ on } S_u\}, \\
  \mathcal{S}_d &= \{d \,|\, d \in H^1(\Omega_1),\; d \in [0,1]\}, \\
  \mathcal{S}_p &= \{\mathbf{p} \,|\, \mathbf{p} \in [H^1(\Omega_3)]^3\}, \\
  \mathcal{S}_q &= \{q \,|\, q \in H^1(\Omega_3)\},
\end{align}
where $\check{\mathbf{u}}$ denotes the prescribed displacement on the Dirichlet boundary $S_u$. Similarly, the sets of admissible test functions are given by
\begin{align}
  \mathcal{V}_u &= \{\delta\mathbf{u} \,|\, \delta\mathbf{u} \in [H^1(\Omega_0)]^{n_{\mathrm{dim}}},\; \delta\mathbf{u} = \mathbf{0} \text{ on } S_u\}, \\
  \mathcal{V}_d &= \{\delta d \,|\, \delta d \in H^1(\Omega_1),\; \delta d \geq 0\}, \\
  \mathcal{V}_p &= \{\delta\mathbf{p} \,|\, \delta\mathbf{p} \in [H^1(\Omega_3)]^3\}, \\
  \mathcal{V}_q &= \{\delta q \,|\, \delta q \in H^1(\Omega_3)\},
\end{align}
where $H^1$ denotes the Sobolev space of degree one. The stress tensors for each subdomain are derived in Appendix~\ref{sec:appendix_stress}.

\subsection{Finite element discretization}\label{sec:discretization}

The fields are discretized using continuous Lagrange finite elements:
\begin{align}
  \mathbf{u}^h(\mathbf{X}) &= \sum_{a=1}^{n_u} \mathbf{N}^u_a(\mathbf{X})\mathbf{u}_a, \qquad
  d^h(\mathbf{X}) = \sum_{a=1}^{n_d} N^d_a(\mathbf{X})d_a, \label{eqn:interp_u_d} \\
  p_i^h(\mathbf{X}) &= \sum_{a=1}^{n_p} N^p_a(\mathbf{X})p_{i,a}, \qquad
  q^h(\mathbf{X}) = \sum_{a=1}^{n_q} N^q_a(\mathbf{X})q_a, \label{eqn:interp_p_q}
\end{align}
where $N_a$ denotes shape functions, subscript $a$ denotes node indices, and $n_u$, $n_d$, $n_p$, $n_q$ are the number of nodes associated with each field.

The displacement field uses linear or quadratic Lagrange elements, while the phase-field and auxiliary fields use linear Lagrange elements.

The discrete residual vectors are:
\begin{align}
  \mathbf{R}^u_a &= \int_{\Omega_0} (\mathbf{B}^u_a)^{\mathrm{T}}\mathbf{P}\,\mathrm{d}V - \mathbf{f}^{\mathrm{ext}}_a, \label{eqn:R_u} \\
  R^d_a &= \int_{\Omega_1}\left[(g'(d)\Psi_1^{+} + \frac{G_c}{c_w\ell}w'(d))N^d_a + \frac{2G_c\ell}{c_w}(\mathbf{B}^d_a)^{\mathrm{T}}\nabla d\right]\mathrm{d}V, \label{eqn:R_d} \\
  R^{p_i}_a &= \int_{\Omega_3}\gamma\left[\beta_1(p_i-W_i)N^p_a + \alpha_r(\mathbf{B}^p_a)^{\mathrm{T}}\nabla p_i\right]\mathrm{d}V, \label{eqn:R_p} \\
  R^q_a &= \int_{\Omega_3}\gamma\left[\beta_2(q-J)N^q_a + \alpha_r(\mathbf{B}^q_a)^{\mathrm{T}}\nabla q\right]\mathrm{d}V, \label{eqn:R_q}
\end{align}
where $\mathbf{B}_a = \nabla N_a$ is the gradient of the shape function.

The consistent tangent matrices are obtained by linearization:
\begin{align}
  \mathbf{K}^{uu}_{ab} &= \int_{\Omega_0} (\mathbf{B}^u_a)^{\mathrm{T}}\mathbb{A}\mathbf{B}^u_b\,\mathrm{d}V, \label{eqn:K_uu} \\
  K^{p_i p_i}_{ab} &= \int_{\Omega_3}\gamma\left[\beta_1 N^p_a N^p_b + \alpha_r(\mathbf{B}^p_a)^{\mathrm{T}}\mathbf{B}^p_b\right]\mathrm{d}V, \label{eqn:K_pp} \\
  K^{qq}_{ab} &= \int_{\Omega_3}\gamma\left[\beta_2 N^q_a N^q_b + \alpha_r(\mathbf{B}^q_a)^{\mathrm{T}}\mathbf{B}^q_b\right]\mathrm{d}V, \label{eqn:K_qq} \\
  \mathbf{K}^{p_i u}_{ab} &= -\int_{\Omega_3}\gamma\beta_1 N^p_a \frac{\partial W_i}{\partial\mathbf{F}}\mathbf{B}^u_b\,\mathrm{d}V, \label{eqn:K_pu} \\
  \mathbf{K}^{qu}_{ab} &= -\int_{\Omega_3}\gamma\beta_2 N^q_a \frac{\partial J}{\partial\mathbf{F}}\mathbf{B}^u_b\,\mathrm{d}V, \label{eqn:K_qu} \\
  K^{dd}_{ab} &= \int_{\Omega_1}\left[(g''(d)\Psi_1^{+} + \frac{G_c}{c_w\ell}w''(d))N^d_aN^d_b + \frac{2G_c\ell}{c_w}(\mathbf{B}^d_a)^{\mathrm{T}}\mathbf{B}^d_b\right]\mathrm{d}V, \label{eqn:K_dd}
\end{align}
where the coupling matrices $\mathbf{K}^{up_i} = (\mathbf{K}^{p_i u})^{\mathrm{T}}$ and $\mathbf{K}^{uq} = (\mathbf{K}^{qu})^{\mathrm{T}}$ follow from symmetry of the tangent operator. Here, $\mathbb{A} = \partial \mathbf{P} / \partial \mathbf{F}$ is the fourth-order elasticity two-point tensor.

\subsection{Nonlinear solution strategy}
The discrete residual vectors for each field are assembled into a global residual vector:
\begin{align}
  \mathbf{R} = [\mathbf{R}^u,\, \mathbf{R}^{\mathbf{p}},\, R^q,\, R^d]^{\mathrm{T}}, \label{eqn:global_residual}
\end{align}
which can be interpreted as the difference between internal and external force vectors:
\begin{align}
  \mathbf{R} = \mathbf{F}_{\mathrm{int}} - \mathbf{F}_{\mathrm{ext}}, \label{eqn:R_int_ext}
\end{align}
where $\mathbf{F}_{\mathrm{int}}$ contains contributions from all internal energy terms and $\mathbf{F}_{\mathrm{ext}}$ represents external loads. The system is in equilibrium when $\mathbf{R} = \mathbf{0}$.

For the linearization of the residual, the consistent tangent stiffness matrix $\mathbf{K}$ is assembled from the block matrices defined in Eqs.~\eqref{eqn:K_uu}--\eqref{eqn:K_dd}. The mechanical/contact subsystem involves the coupled tangent matrix with blocks $\mathbf{K}^{uu}$, $\mathbf{K}^{u\mathbf{p}}$, $\mathbf{K}^{uq}$, $\mathbf{K}^{\mathbf{p}u}$, $\mathbf{K}^{\mathbf{pp}}$, $\mathbf{K}^{qu}$, $K^{qq}$, reflecting the coupling between displacement and auxiliary fields. The global system of equations is then given by
\begin{align}
  \begin{Bmatrix} \mathbf{u} \\ \mathbf{p} \\ q \\ d \end{Bmatrix}_{t+\Delta t}
  = \begin{Bmatrix} \mathbf{u} \\ \mathbf{p} \\ q \\ d \end{Bmatrix}_{t}
  - \begin{bmatrix} \mathbf{K}^{uu} & \mathbf{K}^{u\mathbf{p}} & \mathbf{K}^{uq} & \mathbf{0} \\ \mathbf{K}^{\mathbf{p}u} & \mathbf{K}^{\mathbf{pp}} & \mathbf{0} & \mathbf{0} \\ \mathbf{K}^{qu} & \mathbf{0} & K^{qq} & \mathbf{0} \\ \mathbf{0} & \mathbf{0} & \mathbf{0} & K^{dd} \end{bmatrix}_{t}^{-1}
  \begin{Bmatrix} \mathbf{R}^u \\ \mathbf{R}^{\mathbf{p}} \\ R^q \\ R^d \end{Bmatrix}_{t}, \label{eqn:global_system}
\end{align}
where $t+\Delta t$ denotes the current step and $t$ denotes the previous step, and $\mathbf{K}^{\mathbf{pp}}$ and $\mathbf{K}^{\mathbf{p}u}$ denote the block-diagonal and block-row assemblies over all rotation components $i$.

In monolithic solution schemes, the subsystems for displacement, auxiliary fields, and phase-field are solved simultaneously. In contrast, staggered solution schemes solve each system sequentially. While the monolithic scheme is unconditionally stable, the total potential energy is non-convex with respect to $(\mathbf{u}, \mathbf{p}, q)$ and $d$. On the other hand, in the staggered scheme, the global system of equations is solved by fixing either $(\mathbf{u}, \mathbf{p}, q)$ or $d$, rendering each subproblem convex. A staggered (alternate minimization) scheme is employed to solve the coupled problem. The implementation is based on FEniCS \citep{LoggMardalEtAl2012a,AlnaesBlechta2015a}, and the block Jacobian matrices for the multi-field problem are assembled using multiphenics \citep{ballarin2019multiphenics}, which facilitates the definition of block-structured variational forms and subdomain-restricted fields within the FEniCS framework. At each load step $n \to n+1$, the mechanical/contact subsystem (displacement $\mathbf{u}$ and auxiliary fields $\mathbf{p}$, $q$) is solved monolithically using PETSc/SNES with a Newton method and direct (MUMPS) linear solvers. The damage subproblem for $d$ is a bound-constrained quadratic minimization solved using the variational inequality solver PETSc/TAO with the TRON algorithm \citep{balay2019petsc}. This process repeats until a level of convergence specified within the SNES and TAO solvers is achieved.

\begin{center}
  \begin{minipage}{0.95\linewidth}
    \begin{algorithm}[H]
      \caption{Staggered scheme for coupled contact-fracture problem}\label{alg:staggered}
      \begin{algorithmic}[1]
        \State \textbf{Initialization:} Given $(\mathbf{u}_n, \mathbf{p}_n, q_n)$ and $d_n$. Set lower bound $d_{\mathrm{lb}} \leftarrow d_n$.
        \State \textbf{Update boundary conditions:} Apply prescribed loads/displacements at $t_{n+1}$.
        \For{$k = 1, \ldots, k_{\max}$}
        \State \textbf{Mechanical/contact step:} Solve for $(\mathbf{u}^{(k)}, \mathbf{p}^{(k)}, q^{(k)})$ with $d^{(k-1)}$ fixed (SNES solver).
        \State \textbf{Damage step:} Solve for $d^{(k)}$ with $(\mathbf{u}^{(k)}, \mathbf{p}^{(k)}, q^{(k)})$ fixed, subject to $d^{(k)} \geq d_{\mathrm{lb}}$ (TAO solver).
        \State \textbf{Check convergence:} Compute $\epsilon_d = \|d^{(k)} - d^{(k-1)}\|_{\infty}$.
        \If{$\epsilon_d < \mathrm{tol}_{\mathrm{AM}}$}
        \State \textbf{break}
        \EndIf
        \EndFor
        \State \textbf{Update lower bound:} $d_{\mathrm{lb}} \leftarrow d^{(k)}$ (enforce irreversibility).
        \State \textbf{Store solution:} $(\mathbf{u}_{n+1}, \mathbf{p}_{n+1}, q_{n+1}) \leftarrow (\mathbf{u}^{(k)}, \mathbf{p}^{(k)}, q^{(k)})$ and $d_{n+1} \leftarrow d^{(k)}$.
      \end{algorithmic}
    \end{algorithm}
  \end{minipage}
\end{center}

\section{Numerical examples}\label{sec:examples}

This section presents numerical examples to validate and demonstrate the proposed unified PFF-TMC framework. The first two examples are two-dimensional plane-strain problems, where the auxiliary rotation field reduces to a scalar $p$ (rather than a vector $\mathbf{p}$); the final example demonstrates the three-dimensional extension. In all cases, external body forces and surface tractions are absent ($\mathbf{B} = \mathbf{0}$, $\mathbf{T} = \mathbf{0}$), and loading is applied exclusively through prescribed displacements on portions of the boundary, i.e., displacement-controlled loading. Because prescribed displacements are imposed directly on the indenter (or platen) boundary, these bodies undergo rigid-body motion with no internal deformation ($\mathbf{F} = \mathbf{I}$, $J = 1$), so that $\Psi_2 = 0$ throughout $\Omega_2$. Consequently, the material parameters assigned to $\Omega_2$ have no influence on the solution, and they are therefore omitted from the parameter tables that follow.

\begin{remark}[Numerical regularization]\label{rmk:J_safe}
  To prevent numerical failure when $J \to 0^+$ in the compressed third medium, a regularized Jacobian is employed:
  \begin{align*}
    \widetilde{J} = \max(J, \varepsilon)
  \end{align*}
  The 2D rotation measure is similarly regularized:
  \begin{align*}
    \tan\varphi = \frac{F_{12}-F_{21}}{F_{11}+F_{22}+\varepsilon}.
  \end{align*}
  In the following examples, $\varepsilon = 10^{-8}$ is used.
\end{remark}

\subsection{Third-medium contact benchmark}\label{sec:benchmark_tmc}

\begin{figure}[!htb]
  \centering
  \includegraphics[width=0.9\textwidth]{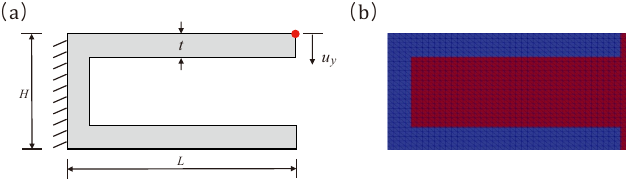}
  \caption{2D C-box benchmark for third-medium contact. (a) Geometry and boundary conditions: the left boundary is clamped, and a prescribed vertical displacement $u_y$ is applied at the top-right corner. (b) Finite element mesh showing the solid domain $\Omega_1$ (blue) and the third-medium domain $\Omega_3$ (red).}
  \label{fig:cbox_setup}
\end{figure}

To validate the third-medium contact component independently, a C-shaped box problem is considered \citep{wriggers2025cma}. A C-shaped solid domain $\Omega_1$ encloses a third-medium cavity $\Omega_3$. The left boundary is fixed ($\mathbf{u} = \mathbf{0}$ on $x = 0$), and a pointwise vertical displacement $u_y$ is applied at the top-right corner. The geometry and finite element mesh are shown in Fig.~\ref{fig:cbox_setup}, and the material parameters are listed in Table~\ref{tab:params_cbox}.

\begin{table}[H]
  \centering
  \caption{Parameters for the 2D C-box benchmark.}
  \label{tab:params_cbox}
  \begin{tabular}{lll}
    \hline
    Parameter & Symbol & Value \\
    \hline
    Shear modulus & $\mu$ & $5/14$ \\
    Bulk modulus & $\kappa$ & $5/3$ \\
    Third-medium scaling & $\gamma$ & $2 \times 10^{-6}$ \\
    Gradient regularization & $\alpha_r$ & $1.0$ \\
    Rotation penalty & $\beta_1$ & $10^4$ \\
    Jacobian penalty & $\beta_2$ & $10$ \\
    \hline
  \end{tabular}
\end{table}

\begin{figure}[!htb]
  \centering
  \includegraphics[width=0.9\textwidth]{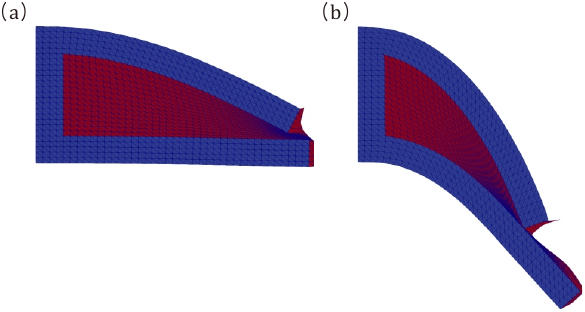}
  \caption{Deformed configurations of the C-box with auxiliary-field regularization: (a) at the onset of contact (corresponding to step 60 in Fig.~\ref{fig:cbox_gap}) and (b) post-contact stage. The solid domain $\Omega_1$ (blue) and the third medium $\Omega_3$ (red) are shown. The auxiliary fields $p$ and $q$ maintain mesh quality in the severely compressed third medium.}
  \label{fig:cbox_result}
\end{figure}

\begin{figure}[!htb]
  \centering
  \includegraphics[width=0.95\textwidth]{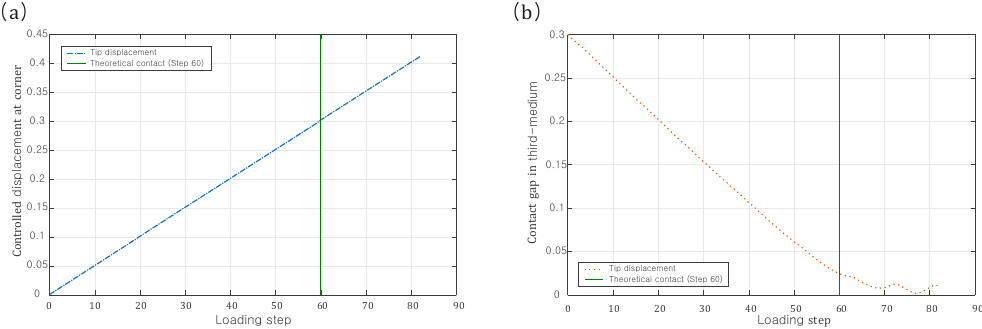}
  \caption{Tip displacement and contact gap as functions of the loading step for the 2D C-box benchmark. The contact gap decreases monotonically until the inner surfaces come into contact through the third medium. The theoretical contact point is indicated by the vertical dashed line.}
  \label{fig:cbox_gap}
\end{figure}

Figure~\ref{fig:cbox_result} shows the deformed configurations with auxiliary-field regularization at two loading stages. As the top arm deflects downward, the gap between the upper and lower arms decreases until the inner surfaces come into contact through the third medium. The auxiliary fields $p$ and $q$ regularize local rotation and volumetric deformation, maintaining mesh quality even under severe compression. Fig.~\ref{fig:cbox_result}(a) shows the configuration at the onset of contact (step 60 in Fig.~\ref{fig:cbox_gap}), and Fig.~\ref{fig:cbox_result}(b) shows the post-contact stage, demonstrating successful self-contact enforcement through the compressed third medium. Figure~\ref{fig:cbox_gap} plots the tip displacement and contact gap as functions of the loading step; the contact gap decreases monotonically until the theoretical contact point (vertical dashed line). As shown in Fig.~\ref{fig:cbox_gap}(b), a residual gap of approximately $0.02$ persists after the theoretical contact point because the third medium, although highly compliant, retains a finite thickness and cannot be compressed to zero volume. It should come as no surprise that exact zero-gap contact cannot be achieved, since the third medium, although highly compliant, retains a finite volume; this is an inherent limitation of the TMC approach, where exact contact is replaced by an asymptotic approximation governed by the parameter $\gamma$.

\begin{figure}[!htb]
  \centering
  \includegraphics[width=0.9\textwidth]{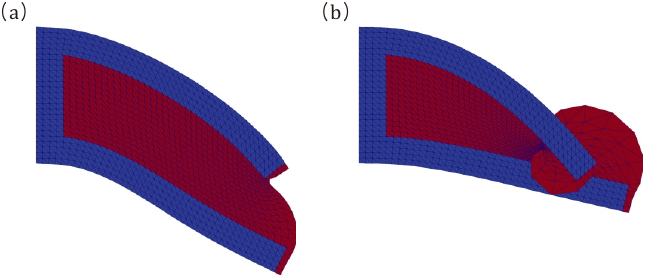}
  \caption{Deformed configurations of the C-box \emph{without} auxiliary-field regularization, demonstrating the necessity of the auxiliary fields $p$ and $q$: (a) the third medium is effectively too stiff, preventing proper contact; (b) the third medium is effectively too soft, leading to excessive deformation and failure to transmit contact forces.}
  \label{fig:cbox_no_reg}
\end{figure}

To further validate the role of the auxiliary-field regularization, the same example is repeated without the auxiliary fields. The auxiliary fields $\mathbf{p}$ and $q$ play an essential role in maintaining well-posed third-medium behavior. Figure~\ref{fig:cbox_no_reg} illustrates the consequences of omitting them: in Fig.~\ref{fig:cbox_no_reg}(a), the effective stiffness is too high and the third medium acts as a rigid barrier preventing proper contact; in Fig.~\ref{fig:cbox_no_reg}(b), it is too low and the third medium undergoes excessive deformation, failing to transmit contact forces. The auxiliary-field regularization resolves this by regularizing the rotation and volumetric fields , as demonstrated in Fig.~\ref{fig:cbox_result}, enabling the third medium to compress uniformly and transmit contact forces in a physically meaningful manner.

\subsection{Phase-field fracture benchmark}\label{sec:benchmark_pff}

To validate the phase-field fracture component independently, a 2D single-edge-notched (SEN) specimen is considered under mode I and mode II loading \citep{miehe2010thermodynamically}. A square plate of side length $L = 1.0$ contains a horizontal edge notch of length $a = L/2$ at mid-height. For mode I loading, the plate is loaded by prescribed vertical displacements:
\begin{align}
  u_y = +\epsilon L \cdot R(t) \quad \text{on } S_{\mathrm{top}}, \qquad
  u_y = -\epsilon L \cdot R(t) \quad \text{on } S_{\mathrm{bot}},
\end{align}
where $R(t)$ is a monotonic ramp from 0 to 1. For mode II loading, a prescribed horizontal displacement is applied to the top boundary while the bottom boundary is fixed. The material and numerical parameters are listed in Table~\ref{tab:params_sen}.

\begin{remark}[Phase-field length scale]\label{rmk:ell}
  The phase-field length scale $\ell$ controls the width of the diffuse crack band and must satisfy $\ell \geq c_{\ell} h_{\min}$ for adequate numerical resolution, where $h_{\min}$ is the minimum element size and $c_{\ell} \geq 1$ is a mesh-dependent constant (typically $c_{\ell} = 1$--$2$). In this work, $\ell = \ell_{\mathrm{multi}} \cdot h_{\min}$ with $\ell_{\mathrm{multi}} \geq 1$ (see Tables~\ref{tab:params_sen} and \ref{tab:params_indentation}).
\end{remark}

\begin{table}[H]
  \centering
  \caption{Parameters for the 2D single-edge-notched (SEN) benchmark.}
  \label{tab:params_sen}
  \begin{tabular}{lll}
    \hline
    Parameter & Symbol & Value \\
    \hline
    Shear modulus & $\mu$ & $20.0$ \\
    Bulk modulus & $\kappa$ & $40.0$ \\
    Fracture toughness & $G_c$ & $0.1$ \\
    Residual stiffness & $k_{\ell}$ & $10^{-6}$ \\
    Length scale multiplier & $\ell_{\mathrm{multi}}$ & $1.0$ \\
    Applied strain (mode I / mode II) & $\epsilon$ & $0.1$ / $0.2$ \\
    Number of load steps (mode I / mode II) & -- & $250$ / $500$ \\
    \hline
  \end{tabular}
\end{table}

\begin{figure}[t]
  \centering
  \includegraphics[width=0.9\textwidth]{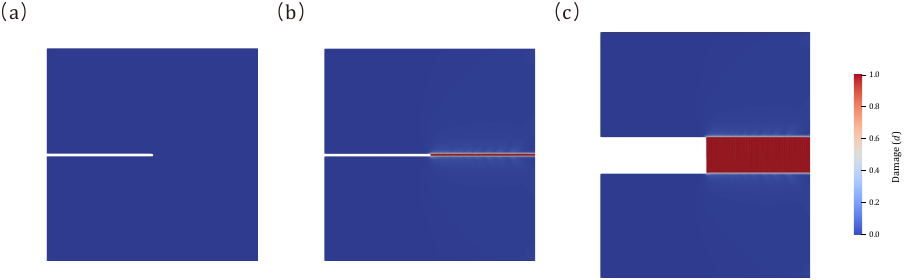}
  \caption{Phase-field fracture benchmark: mode I loading. Damage field $d$ at three stages of crack propagation: (a) initial pre-cracked configuration, (b) damage field in the undeformed (reference) configuration showing horizontal crack growth, and (c) damage field in the deformed (current) configuration.}
  \label{fig:pff_modeI}
\end{figure}

\begin{figure}[t]
  \centering
  \includegraphics[width=0.9\textwidth]{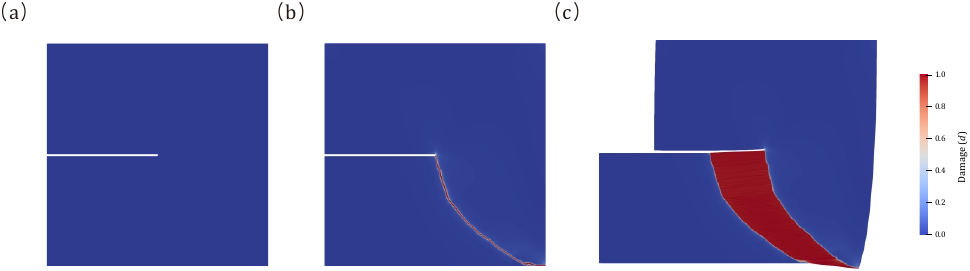}
  \caption{Phase-field fracture benchmark: mode II loading. Damage field $d$ at three stages: (a) initial pre-cracked configuration, (b) damage field in the undeformed (reference) configuration showing crack kinking and subsequent curved propagation, and (c) damage field in the deformed (current) configuration with the characteristic curved crack path under shear loading.}
  \label{fig:pff_modeII}
\end{figure}

Figure~\ref{fig:pff_modeI} shows the damage field for mode I (tensile) loading: Fig.~\ref{fig:pff_modeI}(a) shows the initial pre-notched configuration, while Fig.~\ref{fig:pff_modeI}(b) and Fig.~\ref{fig:pff_modeI}(c) show the final damage field in the reference and deformed configurations, respectively. The crack path is straight and horizontal, consistent with mode I fracture under symmetric tensile loading. Figure~\ref{fig:pff_modeII} shows the corresponding results for mode II (shear) loading. Unlike mode I, the crack kinks away from the initial horizontal direction and propagates along a curved path, as seen in the reference configuration Fig.~\ref{fig:pff_modeII}(b) and the deformed configuration Fig.~\ref{fig:pff_modeII}(c).

\subsection{2D Three-point bending test}\label{sec:three_point_bending}

The unified framework is demonstrated through a three-point bending test, where a rigid circular indenter presses into a pre-cracked beam supported at two points, mediated by a third-medium contact layer. The computational domain is $[0, 2.0] \times [0, 0.7]$, partitioned into the substrate $\Omega_1$ ($y < 0.5$), a circular indenter $\Omega_2$ centered at $(1.0, 0.7)$ with radius $R = 0.19$, and the third medium $\Omega_3$ filling the gap between them. A prescribed downward displacement $u_y = -\check{u}_y$ ($\check{u}_y \in [0, 0.10]$, 500 increments) is applied to the indenter, roller supports ($u_y = 0$) are placed at the bottom corners, and $u_x = 0$ is imposed at $(1.0, 0.5)$ to suppress rigid body motion. The material and numerical parameters are listed in Table~\ref{tab:params_indentation}.

\begin{table}[H]
  \centering
  \caption{Material and numerical parameters for the 2D three-point bending test.}
  \label{tab:params_indentation}
  \begin{tabular}{llll}
    \hline
    Domain & Parameter & Symbol & Value \\
    \hline
    Substrate & Shear modulus & $\mu_1$ & $5/14 \times 10^4$ \\
    & Bulk modulus & $\kappa_1$ & $5/3 \times 10^4$ \\
    & Fracture toughness & $G_c$ & $0.1$ \\
    & Residual stiffness & $k_{\ell}$ & $10^{-6}$ \\
    & Length scale multiplier & $\ell_{\mathrm{multi}}$ & $1.0$ \\
    \hline
    Third medium & Shear modulus & $\mu_3$ & $5/14$ \\
    & Bulk modulus & $\kappa_3$ & $5/3$ \\
    & Scaling parameter & $\gamma$ & $2 \times 10^{-9}$ \\
    & Gradient regularization & $\alpha_r$ & $1.0$ \\
    & Rotation penalty & $\beta_1$ & $10^4$ \\
    & Jacobian penalty & $\beta_2$ & $10$ \\
    \hline
  \end{tabular}
\end{table}

\begin{figure}[!htb]
  \centering
  \includegraphics[width=0.9\textwidth]{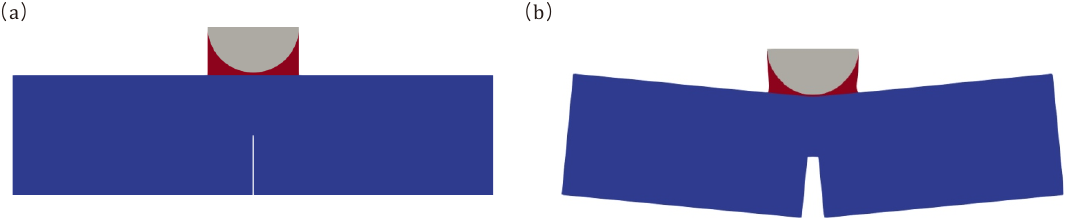}
  \caption{Three-point bending test with rigid indenter: (a) initial configuration showing the substrate $\Omega_1$ (blue), third medium $\Omega_3$ (red), indenter $\Omega_2$ (gray), and vertical pre-crack; (b) deformed configuration showing contact-driven crack opening beneath the indenter.}
  \label{fig:indent_circular}
\end{figure}

Figure~\ref{fig:indent_circular} shows the initial and deformed configurations with the rigid indenter. In Fig.~\ref{fig:indent_circular}(a), a vertical pre-crack is introduced in the substrate beneath the indenter, and the third medium (red) fills the gap between the indenter (gray) and the substrate (blue). In Fig.~\ref{fig:indent_circular}(b), the indenter displaces downward, and the contact condition evolves from an initial point contact to a finite contact area as the indenter conforms to the substrate surface. The third medium naturally accommodates this progressive growth of the contact area, transmitting spatially distributed contact forces (Neumann boundary conditions) to the substrate without requiring explicit contact detection or predefined load distributions. This evolving contact drives the pre-crack to open and propagate.

\begin{figure}[!htb]
  \centering
  \includegraphics[width=0.9\textwidth]{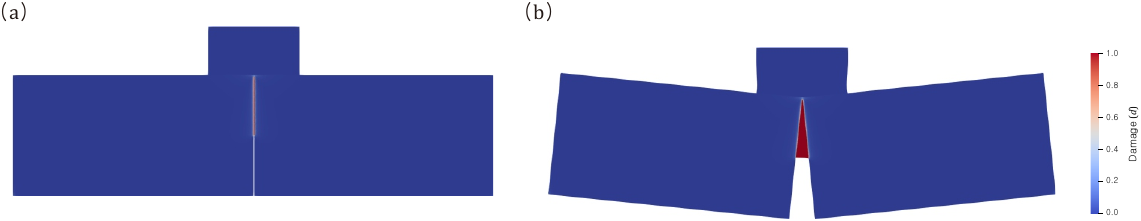}
  \caption{Three-point bending test: phase-field damage evolution. The color scale indicates the damage field $d$, ranging from $d=0$ (blue, intact) to $d=1$ (red, fully broken). (a) Damage field in the reference configuration, showing vertical crack propagation from the pre-crack tip. (b) Damage field in the deformed configuration, displaying the fully developed crack under contact-driven bending.}
  \label{fig:indent_fracture}
\end{figure}

Figure~\ref{fig:indent_fracture} presents the phase-field damage evolution in the three-point bending test. Fig.~\ref{fig:indent_fracture}(a) shows the damage field in the reference configuration with the vertical crack propagating upward from the pre-crack tip beneath the indenter. Fig.~\ref{fig:indent_fracture}(b) shows the damage field in the deformed configuration, displaying the fully developed crack with the beam splitting under the contact-induced bending stress.

\subsection{3D Brazilian disk test}\label{sec:brazilian}

To demonstrate the three-dimensional extension of the unified framework, a Brazilian disk test is considered, a standard method for determining the indirect tensile strength of brittle materials. A cylindrical specimen is compressed diametrally between two loading platens; the resulting stress state induces tensile stresses along the vertical loading diameter, leading to mode~I crack initiation and propagation.

\begin{figure}[!htb]
  \centering
  \includegraphics[width=0.9\textwidth]{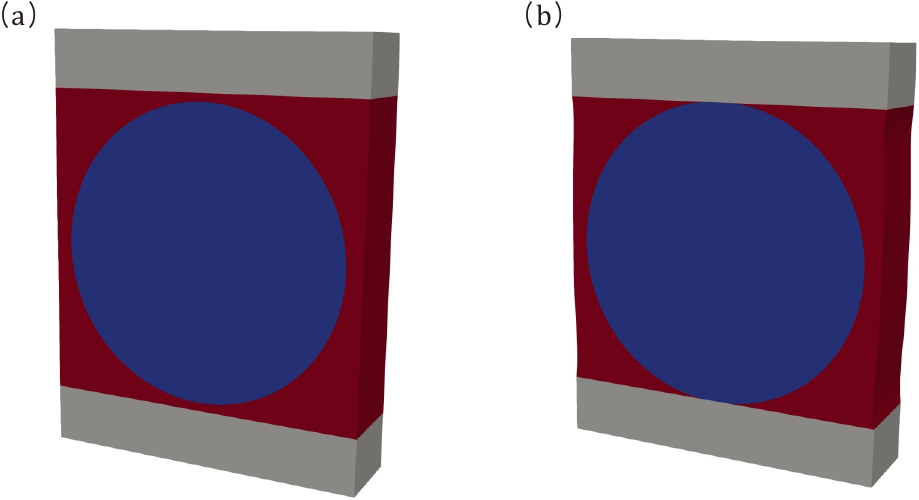}
  \caption{Brazilian disk test: domain decomposition. (a) Undeformed configuration showing the specimen $\Omega_1$ (blue), jaw platens $\Omega_2$ (gray), and third medium $\Omega_3$ (red). (b) Deformed configuration under diametral compression, with the third medium transmitting contact forces from the platens.}
  \label{fig:brazilian_subdomains}
\end{figure}

The computational domain is a three-dimensional extrusion of the two-dimensional cross-section shown in Fig.~\ref{fig:brazilian_subdomains}(a). The specimen $\Omega_1$ is a circular disk of radius $R_1 = 0.90$, the loading platens $\Omega_2$ are placed at the top and bottom of the specimen with a contact gap of $\delta = 0.05$, and the third medium $\Omega_3$ fills the remaining space between the platens and the specimen. The domain extends $[-1.0,1.0]\times[-1.3,1.3]$ in the $xy$-plane and is extruded in the $z$-direction with thickness $T = 0.50$ and three structured layers. Symmetry boundary conditions $u_z = 0$ are imposed on both $z$-faces to enforce a plane-strain-like constraint. Loading is applied by prescribing equal and opposite vertical displacements $u_y = \pm\Gamma/2$ on the top and bottom faces of the platens, with $\Gamma \in [0, 0.3]$ over 300 increments. An initial crack of total length $L_{\mathrm{crack}} = 0.50$ is introduced along the vertical diameter as a phase-field damage profile. The material and numerical parameters are listed in Table~\ref{tab:params_brazilian}.

\begin{table}[H]
  \centering
  \caption{Material and numerical parameters for the 3D Brazilian disk test.}
  \label{tab:params_brazilian}
  \begin{tabular}{llll}
    \hline
    Domain & Parameter & Symbol & Value \\
    \hline
    Specimen & Shear modulus & $\mu_1$ & $5/14 \times 10^5$ \\
    & Bulk modulus & $\kappa_1$ & $5/3 \times 10^5$ \\
    & Fracture toughness & $G_c$ & $0.1$ \\
    & Residual stiffness & $k_{\ell}$ & $10^{-6}$ \\
    & Length scale & $\ell$ & $0.02$ \\
    \hline
    Third medium & Shear modulus & $\mu_3$ & $5/14$ \\
    & Bulk modulus & $\kappa_3$ & $5/3$ \\
    & Scaling parameter & $\gamma$ & $2 \times 10^{-9}$ \\
    & Gradient regularization & $\alpha_r$ & $1.0$ \\
    & Rotation penalty & $\beta_1$ & $10^4$ \\
    & Jacobian penalty & $\beta_2$ & $10$ \\
    \hline
  \end{tabular}
\end{table}

The specimen energy uses the spectral decomposition (Eqs.~\eqref{eqn:Psi_plus}--\eqref{eqn:Psi_minus}), where the three-dimensional principal strains are computed analytically using Cardano's formula with smooth regularization to ensure well-defined Jacobians (see Remark~\ref{rmk:eigenvalues}). The jaw platens and third medium retain the finite-deformation Neo-Hookean formulation. The mesh is refined along the expected crack path ($|x| < 0.03$, $|y| < R_1$) with element size $h_{\mathrm{crack}} = 0.005 \approx \ell/4$ to adequately resolve the phase-field crack band, while a base size $h = 0.08$ is used elsewhere.

\begin{figure}[!htb]
  \centering
  \includegraphics[width=0.9\textwidth]{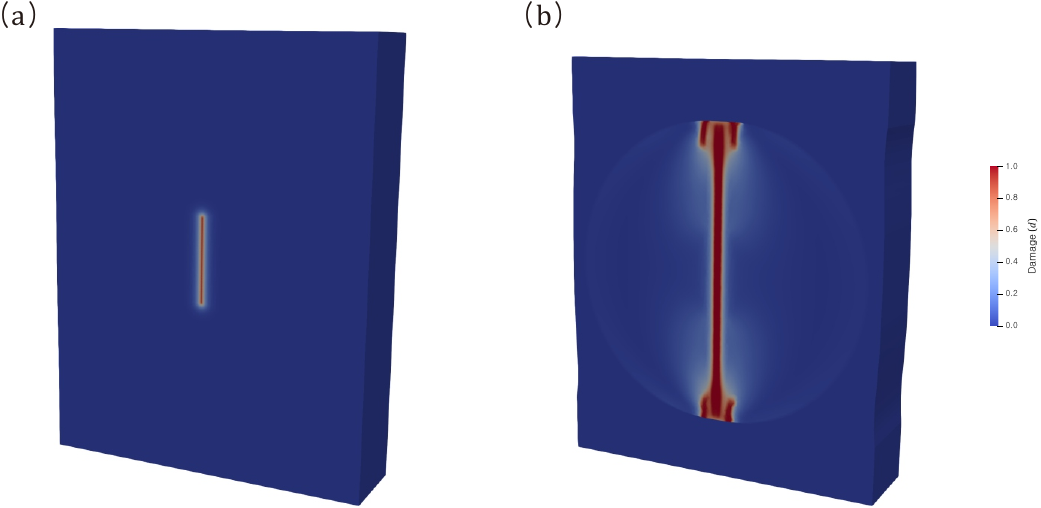}
  \caption{Brazilian disk test: phase-field damage evolution. The color scale indicates the damage field $d$, ranging from $d=0$ (blue, intact) to $d=1$ (red, fully broken). (a) Initial pre-crack configuration with the phase-field damage along the vertical diameter. (b) Fully developed vertical crack after diametral compression, showing the specimen splitting along the loading diameter.}
  \label{fig:brazilian_fracture}
\end{figure}

Figure~\ref{fig:brazilian_subdomains} shows the domain decomposition before and after deformation. Fig.~\ref{fig:brazilian_subdomains}(a) displays the undeformed configuration with the specimen disk $\Omega_1$, the jaw platens $\Omega_2$, and the third medium $\Omega_3$ filling the gap. Fig.~\ref{fig:brazilian_subdomains}(b) shows the deformed configuration with the compressed specimen and the third medium transmitting contact forces from the platens to the specimen surface. Figure~\ref{fig:brazilian_fracture} presents the phase-field damage evolution. Fig.~\ref{fig:brazilian_fracture}(a) shows the initial pre-crack configuration with the phase-field damage along the vertical diameter. In Fig.~\ref{fig:brazilian_fracture}(b), after diametral compression, a fully developed vertical crack splits the specimen along the loading diameter. 

In addition, secondary fracture zones are observed near the contact regions at both ends of the loading diameter, where the evolving contact area between the platens and the specimen produces localized stress concentrations. This result is consistent with experimentally observed fracture patterns in the Brazilian test, where crushing-type damage develops at the loading points \citep{kumar2024strength}. Notably, such contact-induced secondary fracture cannot be reproduced by simplified loading models (e.g., concentrated line loads or prescribed pressure distributions), which impose a fixed contact geometry and thus fail to capture the stress redistribution caused by the progressive growth of the contact area. This demonstrates a key advantage of the TMC approach, which naturally resolves the evolving contact interface and its coupling with damage evolution.

\section{Conclusions}\label{sec:conclusions}

This paper has presented a unified variational framework for phase-field fracture and third-medium contact in finite deformation hyperelasticity. The underlying principle is that both phenomena are treated through regularization: the sharp crack topology is regularized into a diffuse damage field via the phase-field approach, while the discrete contact interface is regularized through a compliant third medium with auxiliary fields. This shared regularization philosophy enables the construction of a single total potential energy functional that seamlessly integrates degraded hyperelastic response in the fracturing substrate, compliant third-medium energy with auxiliary-field regularization, and fracture dissipation through the AT2 crack density functional, without requiring separate algorithmic treatments for contact detection or crack tracking. The coupled problem is solved via a staggered alternate minimization scheme that separates the monolithic mechanical/contact subsystem from the bound-constrained damage minimization, with irreversibility enforced through a monotonically increasing lower bound on the damage field.

After independent validation of the fracture and contact components, the framework has been applied to two coupled contact--fracture problems. In the two-dimensional three-point bending test, the contact condition evolves from an initial point contact to a finite contact area as the rigid indenter conforms to the substrate surface; the third medium naturally resolves this evolving Neumann boundary condition, transmitting spatially distributed contact forces that drive a vertical crack upward from the pre-crack tip. In the three-dimensional Brazilian disk test, the formulation reproduces the characteristic vertical splitting crack along the loading diameter and, notably, secondary fracture zones near the platen--specimen contact regions that arise from the progressive growth of the contact area during compression. Such crushing-type damage at the loading points, consistent with experimental observations, cannot be captured by simplified loading models that prescribe fixed contact geometries. This ability to capture crushing-type secondary fracture at the loading points constitutes a distinctive advantage of the unified TMC-PFF approach: because the third medium naturally resolves the evolving contact interface, the progressive growth of the contact area and the resulting stress redistribution are automatically coupled with damage nucleation, without any algorithmic modifications or a priori assumptions on the contact geometry.

The proposed framework is not without its limitations. First, exact zero-gap contact cannot be achieved: the third medium retains a finite compressed thickness, resulting in a small residual gap whose magnitude depends on the scaling parameter $\gamma$. Reducing $\gamma$ decreases the residual gap but may increase ill-conditioning of the tangent stiffness matrix, requiring a balance between contact accuracy and numerical robustness. Second, the computational cost is substantial, as the third medium introduces additional degrees of freedom and the staggered scheme requires multiple Newton iterations per load step for both the mechanical/contact and damage subsystems. Efficient solution strategies, such as adaptive mesh refinement, parallel solvers, and accelerated staggered schemes, will be essential for extending the framework to large-scale three-dimensional problems.

Several directions for future research emerge from this work. The present work limits itself to employing frictionless contact without adhesion, and extension to Coulomb friction and adhesive contact would require augmenting the third-medium energy with tangential resistance and surface adhesion terms, respectively. Incorporation of inertia effects would enable simulation of impact-induced fracture problems, while extension to anisotropic materials using directional crack density functionals would broaden applicability to fiber-reinforced composites and biological tissues. Finally, integration with thermal effects and chemical processes would address coupled degradation mechanisms relevant to a wider range of engineering systems. The authors plan to apply the proposed framework to nuclear fuel rod analysis, where pellet--cladding mechanical interaction (PCMI), gap conductance evolution, and pellet fragmentation can be simultaneously captured within the unified TMC formulation.

\section*{Acknowledgments}
\begin{itemize}
    \item This research was funded by the `Changwon National University - Samsung Changwon Hospital joint Collaboration Research Support Project' in 2025.
    \item This study was conducted as part of the Glocal University Project, supported by the RISE (Regional Innovation System \& Education) program funded by the Ministry of Education.
\end{itemize}

\section*{Declarations}
The author declares no competing financial interests or personal relationships that could have appeared to influence the work reported in this paper.

\section*{CRediT Author Contributions}
\textbf{Jaemin Kim}: Methodology, Software, Validation, Formal analysis, Investigation, Data curation, Visualization, Writing -- original draft, Writing -- review \& editing, Project administration, Supervision.
\textbf{Gukheon Kim}: Data curation, Visualization.
\textbf{Sungmin Yoon}: Writing -- review \& editing.
\textbf{Dong-Hwa Lee}: Conceptualization, Resources, Funding acquisition, Writing -- review \& editing.

\begin{appendices}

  \section{Derivation of constitutive relations for phase-field damage}\label{sec:appendix_thermo}
  \renewcommand{\thefigure}{A\arabic{figure}}
  \setcounter{figure}{0}
  \renewcommand{\theequation}{A\arabic{equation}}
  \setcounter{equation}{0}

  The second law of thermodynamics, expressed as the Clausius--Duhem inequality for isothermal processes, requires that the internal dissipation rate be non-negative. For a body undergoing quasi-static deformation with evolving damage, the global dissipation inequality reads \citep{wu2017unified,mao2018theory}:
  \begin{align}
    \dot{\mathcal{D}}
    &= \int_{\Omega_1} \mathbf{P}:\dot{\mathbf{F}} \,\mathrm{d}V
    + \int_{\Omega_1} \boldsymbol{\xi}\cdot\nabla\dot{d} \,\mathrm{d}V
    + \int_{\Omega_1} \omega\,\dot{d} \,\mathrm{d}V
    - \int_{\Omega_1}\dot{\Psi}\,\mathrm{d}V
    - \int_{\Gamma} G_{c}\,\mathrm{d}S
    \geq 0. \label{eqn:A_dissipation}
  \end{align}
  Here, $\mathbf{P}$ is the first Piola--Kirchhoff stress (macroscopic stress), $\boldsymbol{\xi}$ is the vector microstress (energetically conjugate to $\nabla d$), and $\omega$ is the scalar microstress (energetically conjugate to $d$).

  Expanding the rate of free energy using the chain rule:
  \begin{align}
    \dot{\Psi} = \frac{\partial \Psi}{\partial \mathbf{F}}:\dot{\mathbf{F}} 
    + \frac{\partial \Psi}{\partial d}\dot{d} 
    + \frac{\partial \Psi}{\partial \nabla d}\cdot\nabla\dot{d}. \label{eqn:A_Psi_rate}
  \end{align}
  Substituting \eqref{eqn:A_Psi_rate} into \eqref{eqn:A_dissipation} and rearranging:
  \begin{align}
    \dot{\mathcal{D}}
    = \int_{\Omega_1} \left(\mathbf{P} - \frac{\partial \Psi}{\partial \mathbf{F}}\right):\dot{\mathbf{F}} \,\mathrm{d}V
    + \int_{\Omega_1} \left(\boldsymbol{\xi} - \frac{\partial \Psi}{\partial \nabla d}\right)\cdot\nabla\dot{d} \,\mathrm{d}V
    + \int_{\Omega_1} \left(\omega - \frac{\partial \Psi}{\partial d}\right)\dot{d} \,\mathrm{d}V
    - \int_{\Gamma} G_{c}\,\mathrm{d}S
    \geq 0. \label{eqn:A_dissipation_rearranged}
  \end{align}

  For the inequality to hold for all admissible processes, and invoking the standard Coleman--Noll argument for the terms multiplying arbitrary rates $\dot{\mathbf{F}}$ and $\nabla\dot{d}$, the constitutive relations follow:
  \begin{align}
    \mathbf{P} &= \frac{\partial \Psi}{\partial \mathbf{F}} = g(d)\frac{\partial \Psi_{\mathrm{und}}(\mathbf{F})}{\partial \mathbf{F}}, \\
    \boldsymbol{\xi} &= \frac{\partial \Psi}{\partial \nabla d} = \frac{2G_c\ell}{c_w}\nabla d, \\
    \omega &= \frac{G_c}{c_w\ell}w'(d) + g'(d)\,\Psi_{\mathrm{und}}(\mathbf{F}).
  \end{align}

  For the strain energy decomposition $\Psi_{\mathrm{und}} = \Psi_1^{+} + \Psi_1^{-}$, where $\Psi_1^{+}$ is the tensile and $\Psi_1^{-}$ is the compressive part, the degraded substrate strain energy becomes:
  \begin{align}
    \Psi_1(\boldsymbol{\varepsilon}, d) = (g(d) + k_{\ell})\,\Psi_1^{+}(\boldsymbol{\varepsilon}) + \Psi_1^{-}(\boldsymbol{\varepsilon}),
  \end{align}
  where only the tensile part is degraded. This decomposition ensures that cracks can open under tension through the degradation of $\Psi_1^{+}$, while fully damaged material can still resist compression through $\Psi_1^{-}$, thereby preventing crack interpenetration.

  \section{Derivation of crack evolution equation}\label{sec:appendix_crack}
  \renewcommand{\thefigure}{B\arabic{figure}}
  \setcounter{figure}{0}
  \renewcommand{\theequation}{B\arabic{equation}}
  \setcounter{equation}{0}

  Starting from the microscopic balance \eqref{eqn:micro_eq}, multiply by an admissible test function $\delta d$ and integrate over $\Omega_1$:
  \begin{equation}\label{eqn:B1}
    \int_{\Omega_1} \left(\omega-\nabla\cdot\boldsymbol{\xi}\right)\delta d\,\mathrm{d}V = 0.
  \end{equation}
  Applying the product rule and divergence theorem gives
  \begin{align}\label{eqn:B2}
    \int_{\Omega_1} \omega\,\delta d\,\mathrm{d}V
    + \int_{\Omega_1} \boldsymbol{\xi}\cdot\nabla\delta d\,\mathrm{d}V
    - \int_{S\setminus S_d} \left(\boldsymbol{\xi}\cdot\mathbf{N}\right)\delta d\,\mathrm{d}A
    = 0.
  \end{align}
  With the natural boundary condition $\boldsymbol{\xi}\cdot\mathbf{N}=0$ on $S\setminus S_d$, the following is obtained:
  \begin{align}\label{eqn:B3}
    \int_{\Omega_1} \omega\,\delta d\,\mathrm{d}V
    + \int_{\Omega_1} \boldsymbol{\xi}\cdot\nabla\delta d\,\mathrm{d}V
    = 0.
  \end{align}
  Substituting the constitutive relations and the AT2 crack density ($w(d)=d^2$, $w'(d)=2d$, $c_w=2$) yields
  \begin{align}\label{eqn:B4}
    \int_{\Omega_1} \left[-2\left(1-d\right)\Psi^{+}\,\delta d
    + \frac{G_{c}}{c_{w}} \left(\frac{2d}{\ell}\,\delta d + 2\ell\,\nabla d\cdot\nabla\delta d\right) \right]\,\mathrm{d}V = 0,
  \end{align}
  which corresponds to the weak form \eqref{eqn:weak_d} for the AT2 model. The strong form is:
  \begin{align}
    f := -2\left(1-d\right)\Psi^{+}
    + \frac{G_{c}}{c_{w}} \left(\frac{2}{\ell}d - 2\ell \nabla\cdot\nabla d\right) = 0 \quad \text{in } \Omega_1. \label{eqn:strong_d}
  \end{align}
  Together with the irreversibility constraint $\dot{d} \geq 0$, the Karush--Kuhn--Tucker (KKT) conditions are:
  \begin{align}
    \dot{d} \geq 0, \qquad f \geq 0, \qquad \dot{d} \cdot f = 0 \quad \text{in } \Omega_1. \label{eqn:KKT}
  \end{align}

  \section{Derivations of stress tensors}\label{sec:appendix_stress}
  \renewcommand{\theequation}{C\arabic{equation}}
  \setcounter{equation}{0}

  \paragraph{Substrate (spectral split)}
  The substrate employs the small-strain spectral decomposition (Section~\ref{sec:substrate_energy_split}). The Cauchy stress tensor is
  \begin{align}
    \boldsymbol{\sigma}_1 = (g(d)+k_{\ell})\frac{\partial\Psi_1^{+}}{\partial\boldsymbol{\varepsilon}} + \frac{\partial\Psi_1^{-}}{\partial\boldsymbol{\varepsilon}},
  \end{align}
  where
  \begin{align}
    \frac{\partial\Psi_1^{\pm}}{\partial\boldsymbol{\varepsilon}} = \lambda\langle\mathrm{tr}\,\boldsymbol{\varepsilon}\rangle_{\pm}\,\mathbf{I} + 2\mu\,\boldsymbol{\varepsilon}^{\pm},
  \end{align}
  with $\boldsymbol{\varepsilon}^{\pm} = \sum_{i}\langle\varepsilon_i\rangle_{\pm}\,\mathbf{n}_i\otimes\mathbf{n}_i$ as defined in Eq.~\eqref{eqn:eps_pm}.

  \paragraph{Third medium (Neo-Hookean with regularization)}
  The first Piola--Kirchhoff stress for the third medium is
  \begin{align}
    \mathbf{P}_3 = \frac{\partial\Psi_{3}^{\mu}}{\partial\mathbf{F}} + \frac{\partial\Psi_{3}^{\gamma}}{\partial\mathbf{F}} + \frac{\partial\Psi_{3}^{\mathrm{reg}}}{\partial\mathbf{F}}.
  \end{align}
  The volumetric contribution is
  \begin{align}
    \mathbf{P}_{\mathrm{vol}} = \frac{\partial\Psi_{\mathrm{vol}}}{\partial\mathbf{F}} = \kappa(\ln J)\mathbf{F}^{-\mathrm{T}},
  \end{align}
  and the isochoric contribution is
  \begin{align}
    \mathbf{P}_{\mathrm{iso}} = \frac{\partial\Psi_{\mathrm{iso}}}{\partial\mathbf{F}} = \mu J^{-2/3}\left(\mathbf{F} - \frac{I_1}{3}\mathbf{F}^{-\mathrm{T}}\right).
  \end{align}
  With the regularized Jacobian, $J$ is replaced by $\widetilde{J}$ in all finite-deformation expressions.

\end{appendices}

\bibliographystyle{elsarticle-harv}
\bibliography{references}

@book{itskov2007tensor,
  title     = {Tensor Algebra and Tensor Analysis for Engineers: With Applications to Continuum Mechanics},
  author    = {Itskov, Mikhail},
  year      = {2007},
  publisher = {Springer},
  address   = {Berlin, Heidelberg}
}

@article{carol1997normal,
  title={Normal/shear cracking model: application to discrete crack analysis},
  author={Carol, Ignacio and Prat, Pere C and L{\'o}pez, Carlos M},
  journal={Journal of engineering mechanics},
  volume={123},
  number={8},
  pages={765--773},
  year={1997},
  publisher={American Society of Civil Engineers}
}

@article{moes1999finite,
  title   = {A finite element method for crack growth without remeshing},
  author  = {Mo\"{e}s, Nicolas and Dolbow, John and Belytschko, Ted},
  journal = {International Journal for Numerical Methods in Engineering},
  volume  = {46},
  number  = {1},
  pages   = {131--150},
  year    = {1999}
}

@article{johnson1982one,
  title={One hundred years of Hertz contact},
  author={Johnson, Kenneth L},
  journal={Proceedings of the Institution of Mechanical Engineers},
  volume={196},
  number={1},
  pages={363--378},
  year={1982},
  publisher={SAGE Publications Sage UK: London, England}
}

@article{belytschko2009review,
  title   = {A review of extended/generalized finite element methods for material modeling},
  author  = {Belytschko, Ted and Gracie, Robert and Ventura, Giulio},
  journal = {Modelling and Simulation in Materials Science and Engineering},
  volume  = {17},
  number  = {4},
  pages   = {043001},
  year    = {2009}
}

@article{xu1994numerical,
  title   = {Numerical simulations of fast crack growth in brittle solids},
  author  = {Xu, Xiao-Ping and Needleman, Alan},
  journal = {Journal of the Mechanics and Physics of Solids},
  volume  = {42},
  number  = {9},
  pages   = {1397--1434},
  year    = {1994}
}

@article{park2011cohesive,
  title   = {Cohesive zone models: a critical review of traction-separation relationships across fracture surfaces},
  author  = {Park, Kyoungsoo and Paulino, Glaucio H.},
  journal = {Applied Mechanics Reviews},
  volume  = {64},
  number  = {6},
  pages   = {060802},
  year    = {2011}
}

@book{wriggers2006computational,
  title     = {Computational Contact Mechanics},
  author    = {Wriggers, Peter},
  year      = {2006},
  edition   = {2nd},
  publisher = {Springer},
  address   = {Berlin, Heidelberg}
}

@article{li2011crack,
  title={Crack pattern formation in thin film lithium-ion battery electrodes},
  author={Li, Juchuan and Dozier, Alan K and Li, Yunchao and Yang, Fuqian and Cheng, Yang-Tse},
  journal={Journal of The Electrochemical Society},
  volume={158},
  number={6},
  pages={A689--A694},
  year={2011},
  publisher={The Electrochemical Society, Inc.}
}

@article{chen2019approaching,
  title     = {Approaching practically accessible solid-state batteries: stability issues related to solid electrolytes and interfaces},
  author    = {Chen, Rusong and Li, Qinghao and Yu, Xiqian and Chen, Liquan and Li, Hong},
  journal   = {Chemical reviews},
  volume    = {120},
  number    = {14},
  pages     = {6820--6877},
  year      = {2019},
  publisher = {ACS Publications}
}

@article{di2014new,
  title     = {A new methodology for characterizing traction-separation relations for interfacial delamination of thermal barrier coatings},
  author    = {Di Leo, Claudio V and Luk-Cyr, Jacques and Liu, Haowen and Loeffel, Kaspar and Al-Athel, Khaled and Anand, Lallit},
  journal   = {Acta Materialia},
  volume    = {71},
  pages     = {306--318},
  year      = {2014},
  publisher = {Elsevier}
}

@article{jiang2018numerical,
  title     = {Numerical analyses of the residual stress and top coat cracking behavior in thermal barrier coatings under cyclic thermal loading},
  author    = {Jiang, Jishen and Wang, Weizhe and Zhao, Xiaofeng and Liu, Yingzheng and Cao, Zhaomin and Xiao, Ping},
  journal   = {Engineering Fracture Mechanics},
  volume    = {196},
  pages     = {191--205},
  year      = {2018},
  publisher = {Elsevier}
}

@article{peng2019investigation,
  title     = {Investigation of the fracture behaviors of windshield laminated glass used in high-speed trains},
  author    = {Peng, Yong and Ma, Wen and Wang, Shiming and Wang, Kui and Gao, Guangjun},
  journal   = {Composite Structures},
  volume    = {207},
  pages     = {29--40},
  year      = {2019},
  publisher = {Elsevier}
}

@article{gultekin2016phase,
  title     = {A phase-field approach to model fracture of arterial walls: theory and finite element analysis},
  author    = {G{\"u}ltekin, Osman and Dal, H{\"u}sn{\"u} and Holzapfel, Gerhard A},
  journal   = {Computer methods in applied mechanics and engineering},
  volume    = {312},
  pages     = {542--566},
  year      = {2016},
  publisher = {Elsevier}
}

@article{fereidoonnezhad2017mechanobiological,
  title     = {A mechanobiological model for damage-induced growth in arterial tissue with application to in-stent restenosis},
  author    = {Fereidoonnezhad, B and Naghdabadi, R and Sohrabpour, S and Holzapfel, GA},
  journal   = {Journal of the Mechanics and Physics of Solids},
  volume    = {101},
  pages     = {311--327},
  year      = {2017},
  publisher = {Elsevier}
}

@article{harandi2023numerical,
  title     = {Numerical and experimental studies on crack nucleation and propagation in thin films},
  author    = {Harandi, Ali and Rezaei, Shahed and Aghda, Soheil Karimi and Du, Chaowei and Brepols, Tim and Dehm, Gerhard and Schneider, Jochen M and Reese, Stefanie},
  journal   = {International Journal of Mechanical Sciences},
  volume    = {258},
  pages     = {108568},
  year      = {2023},
  publisher = {Elsevier}
}

@article{dalbosco2024multiscale,
  title     = {Multiscale computational modeling of arterial micromechanics: A review},
  author    = {Dalbosco, Misael and Fancello, Eduardo A and Holzapfel, Gerhard A},
  journal   = {Computer Methods in Applied Mechanics and Engineering},
  volume    = {425},
  pages     = {116916},
  year      = {2024},
  publisher = {Elsevier}
}

@article{saha2002effects,
  title     = {Effects of the substrate on the determination of thin film mechanical properties by nanoindentation},
  author    = {Saha, Ranjana and Nix, William D},
  journal   = {Acta materialia},
  volume    = {50},
  number    = {1},
  pages     = {23--38},
  year      = {2002},
  publisher = {Elsevier}
}

@article{hay2011measuring,
  title     = {Measuring substrate-independent modulus of thin films},
  author    = {Hay, Jennifer and Crawford, Bryan},
  journal   = {Journal of Materials Research},
  volume    = {26},
  number    = {6},
  pages     = {727--738},
  year      = {2011},
  publisher = {Cambridge University Press}
}

@article{baldelli2014variational,
  title     = {A variational model for fracture and debonding of thin films under in-plane loadings},
  author    = {Baldelli, AA Le{\'o}n and Babadjian, J-F and Bourdin, Blaise and Henao, Duvan and Maurini, Corrado},
  journal   = {Journal of the Mechanics and Physics of Solids},
  volume    = {70},
  pages     = {320--348},
  year      = {2014},
  publisher = {Elsevier}
}

@article{guillen2019fracture,
  title     = {Fracture analysis of thin films on compliant substrates: A numerical study using the phase field approach of fracture},
  author    = {Guill{\'e}n-Hern{\'a}ndez, Teresa and Reinoso, Jose' and Paggi, M},
  journal   = {International Journal of Pressure Vessels and Piping},
  volume    = {175},
  pages     = {103913},
  year      = {2019},
  publisher = {Elsevier}
}

@article{li2024phase,
  title     = {A phase field fracture model for ultra-thin micro-/nano-films with surface effects},
  author    = {Li, Peidong and Li, Weidong and Tan, Yu and Fan, Haidong and Wang, Qingyuan},
  journal   = {International Journal of Engineering Science},
  volume    = {195},
  pages     = {104004},
  year      = {2024},
  publisher = {Elsevier}
}

@article{francfort1998revisiting,
  title     = {Revisiting brittle fracture as an energy minimization problem},
  author    = {Francfort, Gilles A and Marigo, J-J},
  journal   = {Journal of the Mechanics and Physics of Solids},
  volume    = {46},
  number    = {8},
  pages     = {1319--1342},
  year      = {1998},
  publisher = {Elsevier}
}

@article{miehe2010thermodynamically,
  title     = {Thermodynamically consistent phase-field models of fracture: Variational principles and multi-field FE implementations},
  author    = {Miehe, Christian and Welschinger, Fabian and Hofacker, Martina},
  journal   = {International journal for numerical methods in engineering},
  volume    = {83},
  number    = {10},
  pages     = {1273--1311},
  year      = {2010},
  publisher = {Wiley Online Library}
}

@article{borden2012phase,
  title     = {A phase-field description of dynamic brittle fracture},
  author    = {Borden, Michael J and Verhoosel, Clemens V and Scott, Michael A and Hughes, Thomas JR and Landis, Chad M},
  journal   = {Computer Methods in Applied Mechanics and Engineering},
  volume    = {217},
  pages     = {77--95},
  year      = {2012},
  publisher = {Elsevier}
}

@article{kristensen2021assessment,
  title={An assessment of phase field fracture: crack initiation and growth},
  author={Kristensen, Philip K and Niordson, Christian F and Mart{\'\i}nez-Pa{\~n}eda, Emilio},
  journal={Philosophical Transactions of the Royal Society A: Mathematical, Physical and Engineering Sciences},
  volume={379},
  number={2203},
  year={2021},
  publisher={The Royal Society}
}

@article{wu2017unified,
  title     = {A unified phase-field theory for the mechanics of damage and quasi-brittle failure},
  author    = {Wu, Jian-Ying},
  journal   = {Journal of the Mechanics and Physics of Solids},
  volume    = {103},
  pages     = {72--99},
  year      = {2017},
  publisher = {Elsevier}
}

@article{vajari2023investigation,
  title     = {Investigation of driving forces in a phase field approach to mixed mode fracture of concrete},
  author    = {Vajari, Sina Abrari and Neuner, Matthias and Arunachala, Prajwal Kammardi and Linder, Christian},
  journal   = {Computer Methods in Applied Mechanics and Engineering},
  volume    = {417},
  pages     = {116404},
  year      = {2023},
  publisher = {Elsevier}
}

@article{pranavi2024unifying,
  title     = {A unifying finite strain modeling framework for anisotropic mixed-mode fracture in soft materials},
  author    = {Pranavi, D and Steinmann, P and Rajagopal, A},
  journal   = {Computational Mechanics},
  volume    = {73},
  number    = {1},
  pages     = {123--137},
  year      = {2024},
  publisher = {Springer}
}

@article{mao2018theory,
  title     = {A theory for fracture of polymeric gels},
  author    = {Mao, Yunwei and Anand, Lallit},
  journal   = {Journal of the Mechanics and Physics of Solids},
  volume    = {115},
  pages     = {30--53},
  year      = {2018},
  publisher = {Elsevier}
}

@article{tang2019phase,
  title     = {Phase field modeling of fracture in nonlinearly elastic solids via energy decomposition},
  author    = {Tang, Shan and Zhang, Gang and Guo, Tian Fu and Guo, Xu and Liu, Wing Kam},
  journal   = {Computer Methods in Applied Mechanics and Engineering},
  volume    = {347},
  pages     = {477--494},
  year      = {2019},
  publisher = {Elsevier}
}

@article{ye2020large,
  title     = {Large strained fracture of nearly incompressible hyperelastic materials: enhanced assumed strain methods and energy decomposition},
  author    = {Ye, Jia-Yu and Zhang, Lu-Wen and Reddy, JN},
  journal   = {Journal of the Mechanics and Physics of Solids},
  volume    = {139},
  pages     = {103939},
  year      = {2020},
  publisher = {Elsevier}
}

@book{LoggMardalEtAl2012a,
  title     = {Automated Solution of Differential Equations by the Finite Element Method},
  author    = {Anders Logg and Kent-Andre Mardal and Garth N. Wells},
  year      = {2012},
  publisher = {Springer}
}

@article{AlnaesBlechta2015a,
  title   = {The {FEniCS} Project Version 1.5},
  author  = {Martin S. Aln{\ae}s and Jan Blechta and Johan Hake and August Johansson and Benjamin Kehlet and Anders Logg and Chris Richardson and Johannes Ring and Marie E. Rognes and Garth N. Wells},
  year    = {2015},
  journal = {Archive of Numerical Software},
  volume  = {3},
  number  = {100},
  page    = {9-23}
}

@misc{ballarin2019multiphenics,
  title  = {multiphenics -- easy prototyping of multiphysics problems in {FEniCS}},
  author = {Ballarin, Francesco},
  year   = {2019},
  url    = {https://github.com/multiphenics/multiphenics}
}

@article{kim2025film,
  title     = {A phase-field fracture model for 3D film-substrate systems},
  author    = {Kim, San and Kim, Jaemin},
  journal   = {International Journal of Non-Linear Mechanics},
  volume    = {175},
  pages     = {105126},
  year      = {2025},
  publisher = {Elsevier}
}

@article{pranavi2024anisotropic,
  title     = {Phase field modeling of anisotropic fracture},
  author    = {Pranavi, D and Rajagopal, A and Reddy, J N},
  journal   = {Continuum Mechanics and Thermodynamics},
  volume    = {36},
  pages     = {1267--1282},
  year      = {2024},
  publisher = {Springer}
}

@article{pillai2024length,
  title     = {A phase-field length scale insensitive mode-dependent fracture model for brittle failure},
  author    = {Unnikrishna Pillai, Ayyappan and Behera, Akash Kumar and Rahaman, Mohammad Masiur},
  journal   = {Engineering Fracture Mechanics},
  volume    = {309},
  pages     = {110385},
  year      = {2024},
  publisher = {Elsevier}
}

@article{wu2020anisotropic,
  title     = {A variationally consistent phase-field anisotropic damage model for fracture},
  author    = {Wu, Jian-Ying and Nguyen, Vinh Phu and Zhou, Hao and Huang, Yuli},
  journal   = {Computer Methods in Applied Mechanics and Engineering},
  volume    = {358},
  pages     = {112629},
  year      = {2020},
  publisher = {Elsevier}
}

@article{xue2024egd,
  title     = {An extended gradient damage model for anisotropic fracture},
  author    = {Xue, Liang and Feng, Ye and Ren, Xiaodan},
  journal   = {International Journal of Plasticity},
  volume    = {179},
  pages     = {104042},
  year      = {2024},
  publisher = {Elsevier}
}

@article{balay2019petsc,
  title     = {{PETSc} users manual},
  author    = {Balay, Satish and Abhyankar, Shrirang and Adams, Mark and Brown, Jed and Brune, Peter and Buschelman, Kris and Dalcin, Lisandro and Dener, Alp and Eijkhout, Victor and Gropp, W and others},
  year      = {2019},
  publisher = {Argonne National Laboratory}
}

@article{xiang2007measuring,
  title={Measuring the fracture toughness of ultra-thin films with application to AlTa coatings},
  author={Xiang, Yong and McKinnell, James and Ang, Wie-Ming and Vlassak, Joost J},
  journal={International journal of fracture},
  volume={144},
  number={3},
  pages={173--179},
  year={2007},
  publisher={Springer}
}

@article{amor2009regularized,
  title     = {Regularized formulation of the variational brittle fracture with unilateral contact: Numerical experiments},
  author    = {Amor, Hanen and Marigo, Jean-Jacques and Maurini, Corrado},
  journal   = {Journal of the Mechanics and Physics of Solids},
  volume    = {57},
  number    = {8},
  pages     = {1209--1229},
  year      = {2009},
  publisher = {Elsevier}
}

@article{ambrosio1990approximation,
  title     = {Approximation of functionals depending on jumps by elliptic functionals via $\Gamma$-convergence},
  author    = {Ambrosio, Luigi and Tortorelli, Vincenzo M.},
  journal   = {Communications on Pure and Applied Mathematics},
  volume    = {43},
  number    = {8},
  pages     = {999--1036},
  year      = {1990},
  publisher = {Wiley}
}

@article{bourdin2000numerical,
  title     = {Numerical experiments in revisited brittle fracture},
  author    = {Bourdin, Blaise and Francfort, Gilles A. and Marigo, J.-J.},
  journal   = {Journal of the Mechanics and Physics of Solids},
  volume    = {48},
  number    = {4},
  pages     = {797--826},
  year      = {2000},
  publisher = {Elsevier}
}

@article{pham2011gradient,
  title={Gradient damage models and their use to approximate brittle fracture},
  author={Pham, Kim and Amor, Hanen and Marigo, Jean-Jacques and Maurini, Corrado},
  journal={International Journal of Damage Mechanics},
  volume={20},
  number={4},
  pages={618--652},
  year={2011},
  publisher={SAGE Publications Sage UK: London, England}
}

@article{wriggers2025cma,
  title   = {A third medium approach for contact using first and second order finite elements},
  author  = {Wriggers, Peter and Korelc, Joze and Junker, Philipp},
  journal = {Computer Methods in Applied Mechanics and Engineering},
  volume  = {436},
  pages   = {117740},
  year    = {2025}
}

@article{wriggers2025first,
  title={First order finite element formulations for third medium contact},
  author={Wriggers, Peter and Korelc, Jo{\v{z}}e and Junker, Ph},
  journal={Computational Mechanics},
  volume={76},
  number={3},
  pages={829--845},
  year={2025},
  publisher={Springer}
}

@article{wriggers2026third,
  title={A third medium approach for thermo-mechanical contact based on low order ansatz spaces},
  author={Wriggers, Peter},
  journal={Finite Elements in Analysis and Design},
  volume={255},
  pages={104522},
  year={2026},
  publisher={Elsevier}
}

@misc{dahlberg2025rotation,
  title  = {A rotation-based approach to third medium contact regularization},
  author = {Dahlberg, Vilmer and Sjovall, Filip and Dalklint, Anna and Wallin, Mathias},
  year   = {2025},
  note   = {SSRN preprint, 5776651}
}

@misc{xu2025three,
  title         = {Three-dimensional third medium contact model for hyperelastic contact and pneumatically actuated systems},
  author        = {Xu, Bing-Bing and Xue, Tianju and Wriggers, Peter},
  year          = {2025},
  eprint        = {2512.03181},
  archiveprefix = {arXiv},
  primaryclass  = {math-ph}
}

@article{coleman1963thermodynamics,
  title     = {Thermodynamics with internal state variables},
  author    = {Coleman, Bernard D and Gurtin, Morton E},
  journal   = {The Journal of Chemical Physics},
  volume    = {47},
  number    = {2},
  pages     = {597--613},
  year      = {1967},
  publisher = {American Institute of Physics}
}

@article{kim2025vivo,
  title     = {In vivo deformation of the human basilar artery},
  author    = {Kim, Jaemin and Zhang, Kaiyu and Canton, Gador and Balu, Niranjan and Meyer, Kenneth and Saber, Reza and Paydarfar, David and Yuan, Chun and Sacks, Michael S.},
  journal   = {Annals of Biomedical Engineering},
  volume    = {53},
  number    = {1},
  pages     = {83--98},
  year      = {2025},
  publisher = {Springer}
}

@article{boo2025multiphysics,
  title={Multiphysics modeling of surface diffusion coupled with large deformation in 3D solids},
  author={Boo, Seung-Hwan and Kim, Jaemin},
  journal={European Journal of Mechanics-A/Solids},
  volume={113},
  pages={105713},
  year={2025},
  publisher={Elsevier}
}

@article{ang2022stabilized,
  title={Stabilized formulation for phase-field fracture in nearly incompressible hyperelasticity},
  author={Ang, Ida and Bouklas, Nikolaos and Li, Bin},
  journal={International Journal for Numerical Methods in Engineering},
  volume={123},
  number={19},
  pages={4655--4673},
  year={2022},
  publisher={Wiley Online Library}
}

@article{kim2024finite,
  title     = {A finite element implementation of finite deformation surface and bulk poroelasticity},
  author    = {Kim, Jaemin and Ang, Ida and Ballarin, Francesco and Hui, Chung-Yuen and Bouklas, Nikolaos},
  journal   = {Computational Mechanics},
  volume    = {73},
  number    = {5},
  pages     = {1013--1031},
  year      = {2024},
  publisher = {Springer}
}

@article{kumar2024strength,
  title={The strength of the Brazilian fracture test},
  author={Kumar, Aditya and Liu, Yangyuanchen and Dolbow, John E and Lopez-Pamies, Oscar},
  journal={Journal of the Mechanics and Physics of Solids},
  volume={182},
  pages={105473},
  year={2024},
  publisher={Elsevier}
}

@article{boo2026isogeometric,
  title     = {An isogeometric finite element implementation of visco-hyperelastic {Kirchhoff--Love} shells for large deformation and transient analysis},
  author    = {Boo, Seung-Hwan and Ryu, Hyomin and Kim, Seung Hwan and Kim, Jaemin},
  journal   = {Thin-Walled Structures},
  volume    = {219},
  pages     = {114293},
  year      = {2026},
  publisher = {Elsevier}
}

\end{document}